\documentclass[reprint,superscriptaddress,
nofootinbib,
amsmath,amssymb,aps,
pra,
floatfix,
longbibliography
]{revtex4-1}

\usepackage{graphicx}
\usepackage{dcolumn}
\usepackage{bm}
\usepackage[colorlinks=true,
			linkcolor=blue,
			citecolor=blue,
			anchorcolor=blue,
			urlcolor=blue]{hyperref}
\usepackage{color}
\usepackage{textcomp}
\usepackage{tikz}
\usepackage{units}

\graphicspath{{./},{./figures/}}
\DeclareGraphicsExtensions{.pdf}

\usepackage{times}

\def\barep{\bar \varepsilon}
\def\bara{\bar a}
\def\barb{\bar b}
\def\barl{\bar l}
\def\bark{\bar k}
\newcommand{\br}{\mathbf{r}}
\newcommand{\brp}{{\mathbf r}^{\prime}}
\newcommand{\brpp}{{\mathbf r}^{\prime\prime}}
\newcommand{\dif}{\mathrm{d}}

\newcommand{\UT}{U_\mathrm{T}}
\newcommand{\vf}{v_\mathrm{F}}
\newcommand{\ii}{\rm i}

\begin{document}


\title{Scanning gate microscopy in graphene nanostructures}

\author{Xianzhang Chen}
\affiliation{%
Lanzhou Center for Theoretical Physics and Key Laboratory for Magnetism and Magnetic Materials of MOE, \\
Lanzhou University, Lanzhou, Gansu 730000, China}
\affiliation{Universit\'{e} de Strasbourg, CNRS, Institut de Physique et Chimie des Mat\'{e}riaux de Strasbourg, UMR 7504, F-67000 Strasbourg, France}

\author{Guillaume Weick}
\affiliation{Universit\'{e} de Strasbourg, CNRS, Institut de Physique et Chimie des Mat\'{e}riaux de Strasbourg, UMR 7504, F-67000 Strasbourg, France}

\author{Dietmar Weinmann}
\affiliation{Universit\'{e} de Strasbourg, CNRS, Institut de Physique et Chimie des Mat\'{e}riaux de Strasbourg, UMR 7504, F-67000 Strasbourg, France}

\author{Rodolfo A. Jalabert}
\affiliation{Universit\'{e} de Strasbourg, CNRS, Institut de Physique et Chimie des Mat\'{e}riaux de Strasbourg, UMR 7504, F-67000 Strasbourg, France}


\begin{abstract}
The conductance of graphene nanoribbons and nanoconstrictions under the effect of a scanning gate microscopy tip is systematically studied. Using a scattering approach for noninvasive probes, the first- and second-order conductance corrections caused by the tip potential disturbance are expressed explicitly in terms of the scattering states of the unperturbed structure. Numerical calculations confirm the perturbative results, showing that the second-order term prevails in the conductance plateaus, exhibiting a universal scaling law for armchair graphene strips. For stronger tips, at specific probe potential widths and strengths beyond the perturbative regime, the conductance corrections reveal the appearance of resonances originated from states trapped below the tip. The zero-transverse-energy mode of an armchair metallic strip is shown to be insensitive to the long-range electrostatic potential of the probe.  For nanoconstrictions defined on a strip, scanning gate microscopy allows to get insight into the breakdown of conductance quantization. The first-order correction generically dominates at low tip strength,  while for Fermi energies associated with faint conductance plateaus, the second-order correction becomes dominant for relatively small potential tip strengths. In accordance with the spatial dependence of the partial local density of states, the largest tip effect occurs in the central part of the constriction, close to the edges.  Nanoribbons and nanoconstrictions with zigzag edges exhibit a similar response as in the case of armchair nanostructures,  except when the intervalley coupling induced by the tip potential destroys the chiral edge states. 
\end{abstract}

\maketitle


\section{Introduction}
\label{introduction}

Graphene has established itself as a unique platform for revealing new physical features since it was discovered \cite{Novoselov:2004,Novoselov:2005},  due to the electron pseudorelativistic dispersion and its exceptional high mobilities.  Its two-dimensional carbon atom lattice leads to a gapless linear dispersion at low energies,  and the corresponding electrons,  so-called massless Dirac fermions,  can be described in the low-energy regime by a two-dimensional Dirac equation \cite{Neto:2009}.  Understanding the underlying principles of electronic transport properties of high-mobility graphene nanostructures in the quantum coherent regime is of significant basic interest, as well as crucial for future nanoelectronics applications \cite{Peres:2010}.  A variety of experimental techniques have been employed in this quest, among them, scanning gate microscopy (SGM),  originally applied to the two-dimensional electron gas (2DEG) in the vicinity of a quantum point contact (QPC) defined on a semiconductor heterojunction \cite{Topinka:2000}.

In the SGM technique,  a probe, consisting of a metallized atomic force microscope (AFM) tip, perturbs the sample local electrostatic potential,  while electron transport is measured.  The provided spatial resolution of the tip-position-dependent conductance gives additional information regarding coherent transport,  beyond that gained in a standard transport measurement \cite{sellier2011}.  The 2DEG that forms in the intrinsically two-dimensional structure of graphene is particularly suited for scanning probe techniques.  Several groups have used SGM in order to detect localized charges  \cite{Schnez2010,Pascher2012,Garcia2013} and charge inhomogeneities \cite{Jalilian_2011} in graphene samples. Moreover, SGM techniques have been employed to image 
the electronic cyclotron orbits in graphene \cite{Morikawa2015, Bhandari2016} and scarred wave functions \cite{Cabosart2017}, 
as well as to probe conductance fluctuations \cite{Berezovsky2010a} and weak localization \cite{Berezovsky2010b, Chuang2016} 
for coherent transport in graphene.

An early application to a narrow constriction in a graphene flake \cite{Neubeck_2012} demonstrated that SGM can be used to investigate graphene nanostructures, showing that the conductance is considerably affected  when the tip is at the position of the narrowing. More recent systematic experiments on micrometer-sized constrictions \cite{brun2019,guerra2021} showed that the phenomenon of Klein tunneling considerably modifies the effect of the tip, and can lead to the possibility of Veselago lensing induced by the $n$-$p$-$n$ landscape that appears in the presence of the tip-induced potential.  The SGM has also been used 
to manipulate quantum Hall edge channels \cite{Xiang2016, Bours2017} and
to study the topological breakdown of the quantum Hall effect at the edges of graphene constrictions \cite{Moreau2021, Moreau2022}.

The theoretical study of the Veselago effect in the vicinity of a graphene nanoconstriction (GNC) has been addressed with ray-tracing approaches and tight-binding calculations \cite{guerra2021,Moreau2022}, 
while the magnetic focusing observed with SGM in graphene samples in Refs.~\cite{Morikawa2015, Bhandari2016} has been analyzed 
with the aid of quantum simulations \cite{Petrovic2017}.
Numerical calculations have also been employed in a very thorough investigation of the SGM response in graphene nanoribbons (GNRs) and GNCs with different geometrical features \cite{Mrenca_2015,Mre_ca_Kolasi_ska_2022,Kolasi2017_b}. Within this context, it is important to develop a theoretical approach to the SGM response in order to address the generality of parameter regimes and geometries that can be experimentally encountered.  Therefore,  in this work, we systematically study the effect of a tip-induced potential on the conductance of graphene nanoribbons with and without an additional constriction.  In particular, we generalize the perturbation theory valid in the regime of noninvasive tips \cite{Jalabert2010, Gorini2013} to the situation of graphene and deduce the conductance corrections that appear in low orders of the tip strength. Numerical tight-binding simulations are used to test the limits of these predictions and to address the regime of invasive tips.  

We introduce electronic transport in those systems in Sec.\ \ref{sec:basics} and the description of the tip potential in Sec.~\ref{sec:basics-sgm}. The perturbative approach for low tip strength is presented in Sec.\ \ref{sec:perturbative}, and the results are discussed together with numerically obtained data for armchair edges, including for the nonperturbative regime of invasive tips, in Secs.\ \ref{sec:correction_ribb} and \ref{sec:correction_qpc}. 
The case of zigzag edges is discussed in Sec.~\ref{sec:zigzag}.
After the conclusions (Sec.\ \ref{sec:conclusions}), three appendixes describe technical details of the perturbative approach.

\section{Electronic transport in graphene nanoribbons and nanocontacts}
\label{sec:basics}

\begin{figure}
  \includegraphics[width=\linewidth]{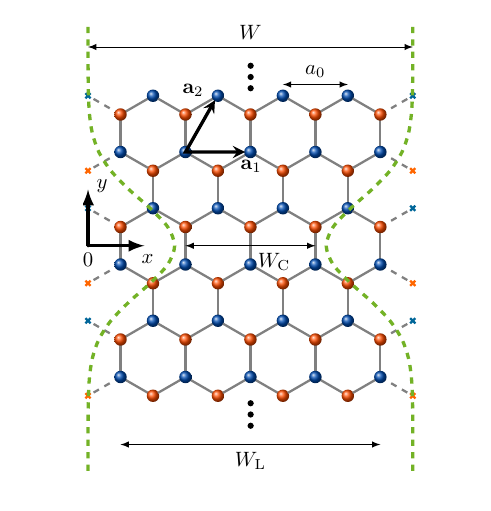}
       \caption{Armchair graphene nanoribbon directed along the vertical ($y$) direction and having a width $W_\mathrm{L}$.  Red and blue dots designate the carbon atoms of the two sublattices within the honeycomb structure. The crosses stand for auxiliary lattice points outside the ribbon where the boundary condition of a vanishing wave function is imposed, and define the effective width $W=W_\mathrm{L}+a_0$. The primitive vectors $\left(\mathbf{a}_1 , \mathbf{a}_2 \right)$ of the hexagonal lattice verify $\left|\mathbf{a}_1\right| =  \left| \mathbf{a}_2 \right| = a_0$.  
The coordinate system $x$-$y$ is chosen such that the Brillouin zone has the two inequivalent corners
$\mathbf{K} = (4\pi/3\, a_0) (1,0)$ and $\mathbf{K}^{\prime} = - \mathbf{K}$.
The graphene nanoconstriction (GNC) geometries are defined by imposing a vanishing wave function outside the green dashed lines according to Eq.~\eqref{eq:QPCdef},  where $W_\mathrm{C}$ is the narrowest width.}
    \label{fig:lattice}
\end{figure}

We choose to characterize the graphene honeycomb structure by the primitive lattice vectors $\mathbf{a}_1 = a_0(1,0)$ and $\mathbf{a}_2 = a_0\left(1/2,\sqrt{3}/2\right)$,  with $a_0=\sqrt{3}\, a_\mathrm{NN}$ and $a_\mathrm{NN} \simeq \unit[0.142]{nm}$ the nearest-neighbor distance. With the convention adopted in Fig.~\ref{fig:lattice}, the two atoms (indicated by blue and red dots) of the conventional cell are located at $(0,0)$ and $a_0\left(0,1/\sqrt{3}\right)$. The associated first Brillouin zone, defined by the reciprocal vectors $\mathbf{b}_1 = (2\pi/a_0) \left(1,-1/\sqrt{3}\right)$ and $\mathbf{b}_2 = (4\pi/\sqrt{3}\,a_0)\left(0,1\right)$, has two inequivalent corners at $\mathbf{K} = (4\pi/3\, a_0) (1,0)$ and $\mathbf{K}^{\prime} = - \mathbf{K}$.   

We describe the electron motion in the $x$-$y$ plane through a nearest-neighbor tight-binding model taking into account the $2p_z$ orbitals of the carbon atoms,  with a hopping constant $\mathtt{t} \simeq \unit[2.7]{eV}$. Within such a model,  low-energy physics is described by a Hamiltonian with two Dirac points located at the $\mathbf{K}$ and $\mathbf{K}^{\prime}$ points of the first Brillouin zone and where the two components of the pseudospinor correspond to the values of the wave function at the two sublattices.  These two valleys (Dirac cones),  with a linear pseudorelativistic dispersion relation and a (Fermi) velocity $v_{\rm F} = \sqrt{3}\, a_0\mathtt{t}/2\hbar \simeq \unit[10^{8}]{cm/s}$,  are independent in bulk graphene, but they are coupled through the boundary conditions in the finite-size samples  that we are interested in.  Within this work we perform numerical calculations using the \textsc{kwant} code \cite{kwant_paper2014}, adopting the tight-binding model to constrained geometries, and we use analytical approaches based on the low-energy description of pseudorelativistic electrons.  

The simplest structure to approach the study of coherent electronic transport in graphene is a quasi-one-dimensional strip,  that is,  a GNR connected to electronic reservoirs (source and drain).  The quest to fabricate atomically precise GNRs  \cite{Lin2008,Jiao2009,Lian2010,2010-Nature-JCai,baringhaus2014,rizzo2020} was initiated soon after the pioneering isolation of graphene monolayers \cite{Novoselov:2004}, while their theoretical characterization in the context of graphite studies \cite{1996-PRB-Nakada}  predated the previous experimental achievements.  In particular,  the electronic and transport properties of perfect GNRs have been shown to depend on the atomic arrangement at the edges of the strip 
\cite{Brey2006,Wurm2009,Yamamoto2009,Wakabayashi2010,Wurm2012,
Bergvall2014,Cao2017},  and to be strongly affected by defects \cite{Munoz:2006,Li2008,Mucciolo2008,Ihnatsenka2009,Orlof2013}.  Considerable progress has been achieved in the fabrication procedures since the diffusive strips that first showed signs of subband quantization formation \cite{Lin2008,Lian2010},  and it is nowadays possible to fabricate very narrow (of the order of $\unit[1]{nm}$ in width) GNRs with atomic precision and specific armchair or zigzag edges \cite{Ruffieux:2016, Kolmer:2020} (although quantum transport measurements seem for now challenging for these very small samples).  For a recent review on GNRs, see Ref.~\cite{2021-NRP-HWang}.

Within the convention of Fig.~\ref{fig:lattice}, cutting the edges of the strip along the $y$ direction gives rise to an armchair GNR,  while cutting along the $x$ axis results in a zigzag GNR (not shown). In the case of an armchair GNR the wave function vanishes on both sublattices at the edges, while for a zigzag GNR the wave function vanishes on a single sublattice at each edge. Denoting $W_\mathrm{L}$ the width of the graphene strip,  we define the effective width $W=W_\mathrm{L}+a_0$ for convenience.  In Appendix \ref{sec:appendixA} we present the electronic eigenstates of armchair nanoribbons, which provide a useful basis for the scattering approach to quantum transport through graphene nanostructures and for the perturbative development of the SGM response.

In our study of quantum transport we consider the linear regime,  where a small applied source-drain bias voltage $V_\mathrm{SD}$ results in a small current $I$, and we focus on the linear conductance $G=I/V_\mathrm{SD}$ at zero temperature. Assuming that the scatterer is connected to two leads that can be assimilated to semi-infinite nanoribbons with $N$ propagating modes labeled by the channel index $a$, the two-terminal Landauer formula reads
\begin{equation}
    \label{eq:conductance}
G =  G_0 \ {\rm Tr}\!\left[t^{\dagger} t  \right]    \,   ,
\end{equation}
where $G_0=2e^2/h$ is the quantum of conductance and $t_{ba}$ the matrix elements of the $N \times N$ transmission amplitude matrix $t$ evaluated at the Fermi energy $E_\mathrm{F}$.

\begin{figure}
   \includegraphics[width=\linewidth]{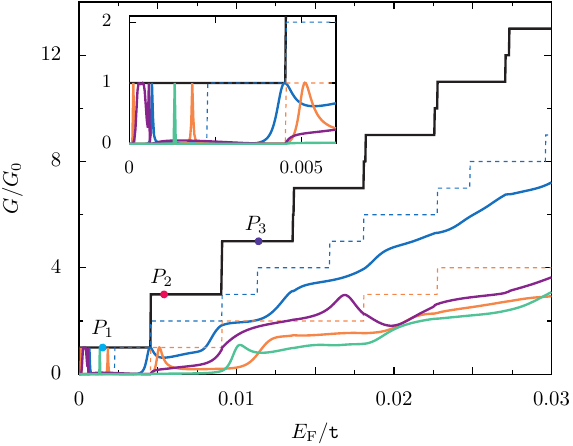}
    \caption{Conductance (in units of the conductance quantum $G_0=2e^2/h$) as a function of the Fermi energy (scaled by the hopping constant $\mathtt{t} = 2\hbar v_\mathrm{F}/\sqrt{3}\,  a_0$) for different armchair GNRs and GNCs. The thick solid black line corresponds to a metallic armchair GNR with a width $W_\mathrm{L}=599\, a_0 \simeq \unit[147.3]{nm}$. The thick colored solid lines represent the conductance of GNCs of different shapes defined by Eq.~\eqref{eq:QPCdef} with $W_\mathrm{C}=399\, a_0$ and $C=\unit[25]{nm}$ (dark blue), $W_\mathrm{C}=199\, a_0$ and $C=\unit[40]{nm}$ (green),  $W_\mathrm{C}=199\, a_0$  and $C=\unit[25]{nm}$ (orange), and $W_\mathrm{C}=199\, a_0$ and $C=\unit[10]{nm}$ (violet). The thin dashed lines stand for the conductance of GNRs with the widths $W_\mathrm{C}$ used to define the corresponding GNC (according to the color convention). Both values of the width $W_\mathrm{C}$ correspond to a semiconductor armchair GNR. The different points $P_i$ indicate the parameters chosen to perform the SGM analysis of GNRs in Sec.~\ref{sec:correction_ribb}.  Inset: Detail of the low-energy sector of the main figure.  
}
\label{fig:TvsE}
\end{figure} 

An infinite GNR provides an example of a perfect scatterer,  with a unitary transmission for all the propagating modes.  Thus,  $G$ is a steplike function of the Fermi energy $E_\mathrm{F}$. In Fig.~\ref{fig:TvsE} we present the numerical results for the conductance of a metallic armchair nanoribbon with a width $W_\mathrm{L}=599\, a_0$ (where electron transport is along the vertical direction $y$) as a function of the Fermi energy (thick black solid line). The zero-transverse-energy mode, not being degenerate, determines the first conductance plateau at $G = G_0$. Upon increasing $E_\mathrm{F}$, the other modes become progressively populated. The Dirac equation predicts these modes to be doubly degenerate, and thus conductance plateaus are separated by steps of $2G_0$. The numerical results agree with such an expectation for small values of $E_\mathrm{F}$, but as $E_\mathrm{F}$ increases, lattice effects become relevant, the mode degeneracy is progressively destroyed, and plateaus with steps of $G_0$ start to appear.   

As commented above,  the conductance quantization obtained in a perfect GNR,  like that of Fig.~\ref{fig:TvsE},  is not robust,  since a moderate amount of edge roughness results in an important suppression of the conductance and the appearance of localized states at the edges 
\cite{Munoz:2006,Li2008,Mucciolo2008,Ihnatsenka2009}.  The sensitivity is most pronounced near the neutrality point,  where the number of propagating modes is small.  By the same token,  the definition of a nanoconstriction in a GNR considerably deteriorates the conductance quantization.  Indeed,  a hallmark of mesoscopic physics, the conductance quantization with plateaus at integer multiples of $G_0$ observed in transport through semiconductor-based QPCs  \cite{vanwees88,wharam88},  is elusive in GNCs,  where no well-defined plateaus are typically observed,  particularly for sharp constrictions.  

The signatures of conductance quantization observed in high-quality suspended \cite{Tombros:2011} or hexagonal-boron-nitride encapsulated \cite{Terres:2016} devices were reproducible modulations (kinks) in the conductance with a spacing that varies in the range of $2$ to $4 e^2/h$.  The differences with the clear quantization characteristic of semiconductor-based QPCs were attributed to strong short-range scattering at the rough edges of the device \cite{Kinikar:2017} favoring the intervalley coupling.  
The use of nanopatterning techniques allowed to define GNCs with high precision and reduced edge roughness,  resulting in a more robust quantization \cite{Kun_2020}.
Depending on the geometry and the smoothness of the edges,  spikes in the conductance might appear due to the resonances associated with the finite length of the constriction \cite{Clerico:2018}. 

Conductance quantization in semiconductor QPCs can be understood from the adiabatic electron transport,  in the case of a smooth constriction with a slow variation in the transverse dimension 
\cite{glazman1998,*glazman1998JETP}, or alternatively,  from the small mismatch between the transmission eigenmodes in the leads and the propagating channels within the junction,  in the case of abrupt geometries \cite{szafer1989}.  Both adiabaticity and small mismatch are problematic in GNCs, resulting in a poor observed conductance quantization.  For the case of zigzag edges,  the adiabatic approximation fails to describe the constriction region, since the electron motion along the stripe is strongly coupled with that in the perpendicular direction,  while for armchair edges the change of edge orientation along the constriction is a source of scattering \cite{Katsnelson:2007}.  Unlike a semiconductor QPC,  the GNC appears as a short-range scatterer,  which, together with other possible defects,  degrades the conductance quantization.  

A million-atom simulation of smooth constrictions resulted in faint quantization steps in integers of $2 e^2/h$,  put in evidence by a clustering of dips in ${\rm d}G/{\rm d} E_\mathrm{F}$ around these conductance values \cite{Ihnatsenka:2012}.  The conductance spikes obtained in geometries like that of  Ref.~\cite{Clerico:2018} were linked to the longitudinal quantization induced by the finite length of the constriction \cite{Yannouleas2015}.

We choose for our model calculations a smooth GNC geometry described by the condition
\begin{equation}
\label{eq:QPCdef}
\left|x-\frac W2\right|<\frac{1}{2}\left[W-(W-W_\mathrm{C})\exp\left(-\frac{y^2}{C^2}\right)
\right]\, ,
\end{equation}
which is illustrated by the green dashed lines in Fig.~\ref{fig:lattice}, where
$W_\mathrm{C}$ is the narrowest width and $C$ characterizes the length of the constriction.  Nonzero values of the electronic wave function are allowed only on lattice sites that fulfill condition \eqref{eq:QPCdef}.  The details of the constriction geometry are known to be determinant for its transport properties.  Depending on the fabrication procedure,  we may have wedgelike constrictions \cite{Terres:2016} or smooth structures \cite{Tombros:2011}.  Our choice \eqref{eq:QPCdef} of a smooth GNC for the SGM calculations is motivated by its simplicity,  and the fact that by varying the two defining parameters $W_\mathrm{C}$ and $C$,  the degree of abruptness can be tuned. 

In agreement with the results of Ref.~\cite{Ihnatsenka:2012},  Fig.~\ref{fig:TvsE} shows that the presence of a constriction reduces the transmission of the ribbon and destroys the conductance plateaus. The thick colored solid lines show the conductance of GNCs of different width ($W_\mathrm{C}$) and length ($C$) of the constriction.  
For each of the two values of $W_\mathrm{C}$ chosen,  the conductance of a GNR with that width is presented by thin dashed lines with the corresponding color, i.e.,  dark blue (orange) for $W_\mathrm{C} = 399\, a_0$ ($199\, a_0$).  These values represent the conductance that would have a perfectly adiabatic constriction characterized by the width $W_\mathrm{C}$.  Since both widths define a semiconductor armchair GNR,  a zero-conductance region appears at low energy,  and for larger $E_\mathrm{F}$ the conductance plateaus are separated by steps of $G_0$.  Only for very short GNCs ($C=\unit[10]{nm}$,  violet line),  tunneling across the constriction permits to exceed,  at specific energies,  the limit set by the corresponding fictitious GNR.  The conductance resonances appearing at low Fermi energy, shown on smaller scale in the inset of Fig.~\ref{fig:TvsE},  have been observed in other numerical simulations \cite{Ihnatsenka:2012,Mrenca_2015,Guimar:2012,Yannouleas2015},  and they can be attributed to quasi-bound states in a GNC \cite{YJXiong:2011,Deng2014}. 

\begin{figure}[tb]
   \includegraphics[width=\linewidth]{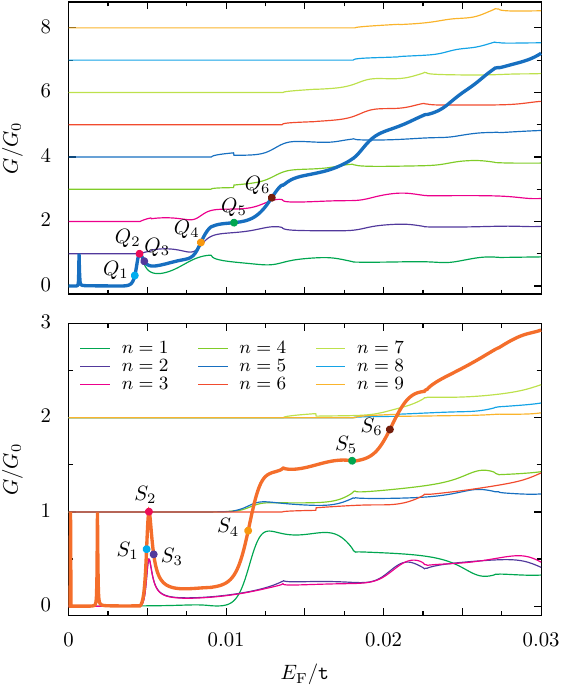}
    \caption{Decomposition of the conductance in units of the conductance quantum (thick curves),  for two of the GNCs  shown in Fig.~\ref{fig:TvsE},  in terms of the transmission eigenmodes (labeled by the index $n$,  thin curves with colors according to the legend) as a function of the Fermi energy (in units of the hopping constant).  Top: $W_\mathrm{C}=399\, a_0$ and $C=\unit[25]{nm}$ (dark blue).  Bottom: $W_\mathrm{C}=199\, a_0$ and $C=\unit[25]{nm}$ (orange).  
For a better visibility, the transmission eigenvalues are ordered and successively shifted by unity in the upper panel, and gathered in three groups, 
each of them shifted by unity, in the lower panel.    
The different points $Q_i$ and $S_i$ mark the parameters chosen to perform the SGM analysis of GNCs in Sec.~\ref{sec:correction_qpc}.
}
\label{fig:TvsEmd}     
\end{figure}

An insight into the origin of the faint conductance plateaus of GNCs can be obtained by working in the representation of the transmission eigenmodes (i.e., the eigenvectors of $t^{\dagger} t$) \cite{jalabert2016},  for which the Landauer formula \eqref{eq:conductance} becomes a diagonal sum,
\begin{equation}
    \label{eq:conductancemd}
G = G_0 \ \sum_{n=1}^{N} \ {\cal T}_n^2    \,   ,
\end{equation}
where the ${\cal T}_n$ are referred to as the transmission eigenvalues associated to the transmission eigenmodes $n$.  In the upper panel of Fig.~\ref{fig:TvsEmd} we reproduce the $E_\mathrm{F}$ dependence of the conductance presented in Fig.~\ref{fig:TvsE} for $W_\mathrm{C}=399\, a_0$ and $C=\unit[25]{nm}$ (thick dark blue line),  decomposed according to the contribution of the transmission eigenvalues (thin colored lines).  We see that as $E_\mathrm{F}$ increases,  different transmission eigenmodes are progressively turned on,  and at some point they approach unitary transmission,  which favors the appearance of faint plateaus with an approximately quantized conductance.

The lower panel of Fig.~\ref{fig:TvsEmd} shows the eigenmode decomposition for the case of another GNC considered in  Fig.~\ref{fig:TvsE},  with $W_\mathrm{C}=199\, a_0$ and $C=\unit[25]{nm}$ (thick orange line),  being more abrupt than the one of the upper panel.  The transmission eigenmodes are also turned on upon increasing $E_\mathrm{F}$,  and when they attain an approximately constant value,  the resulting conductance presents a faint plateau.  However,  the abruptness of this constriction results in transmission eigenvalues with some nonmonotonic dependence on $E_\mathrm{F}$ without achieving the unitary limit. Therefore, the conductance plateaus are less well defined than in the previous case, and farther away from the quantized values.  Another special feature of the second example above is that the $n=2$ and $n=3$ eigenmodes are degenerate up to $E_\mathrm{F}/\mathtt{t}\simeq 0.01$,  with the same transmission eigenvalue.

As we will see in the sequel,  the transmission eigenmode representation greatly helps to understand the SGM response. The latter is strongly dependent on the value of the unperturbed conductance. Parameters used for a detailed study of the SGM response are indicated in Figs.~\ref{fig:TvsE} and \ref{fig:TvsEmd} by the points $P_i$ and $Q_i$ or $S_i$ for GNRs and GNCs, respectively.

\section{SGM in graphene nanostructures}
\label{sec:basics-sgm}

In an SGM setup the electronic conductance is measured while the charged tip of an AFM is scanned over the sample. The electrostatic potential induced by the tip at the level of the two-dimensional graphene electrons can be approximated by a Lorentzian function \cite{Kolasi2017_b,brun2019,brun2020,Moreau2022}, and thus we write the corresponding potential energy as 
\begin{equation}
\label{eq:tip_potential}
U_{\rm T}(\mathbf{r}) = \frac{u_{\rm T}}
{1 + \left(\mathbf{r}-\mathbf{r}_{\rm T}\right)^2/d^2} \,  ,
\end{equation}
where $\mathbf{r}_\mathrm{T}$ stands for the projection of the tip position on the graphene plane, while the potential strength $u_{\rm T}$ and the disturbance width $d$ are related to the voltage applied to the tip and on the distance between the tip and the graphene flake, respectively. Since $d$ is typically much larger than the lattice spacing $a_0$, the perturbing potential $U_\mathrm{T}$ can be taken as a scalar (i.e., the same for both sublattices and without inducing a coupling between them). In experiments, the value of $d$ is limited by the distance of the tip to the graphene sheet, and is then often of the order of $\unit[100]{nm}$ \cite{brun2019}. In order to limit technical issues due to tip potentials extending beyond the system into the leads, we choose smaller values in our theoretical approach, yet at least an order of magnitude larger than $a_0$. 

\begin{figure}
   \includegraphics[width=\linewidth]{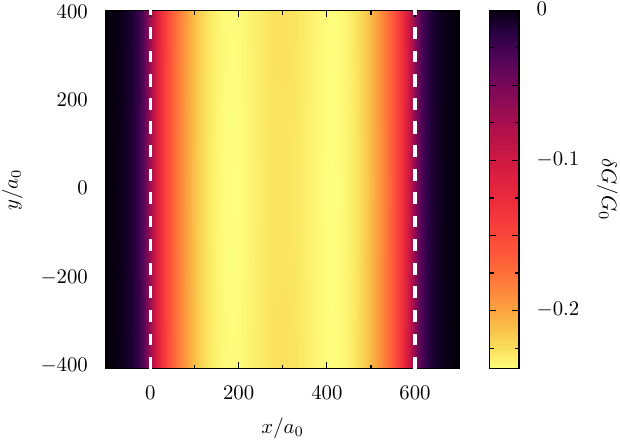} 
    \caption{SGM scan of the conductance correction in color scale for a metallic armchair GNR defined by the white dashed lines ($W=600\, a_0$), for the case where the unperturbed conductance is placed in the second plateau (point $P_2$ of Fig.~\ref{fig:TvsE}). The tip potential is given by Eq.~\eqref{eq:tip_potential},  with $u_{\rm T}=0.02\, \mathtt{t} $ and $d=20\, a_0$.}
    \label{fig:SGMscan_strip}
\end{figure}
As we will show in Sec.~\ref{sec:correction_ribb}, there is no SGM correction for the zero-transverse-energy mode. Therefore, we present in Fig.~\ref{fig:SGMscan_strip} a typical SGM scan,  plotting the numerically obtained  response $\delta G$ (defined as the conductance of the sample with the tip minus the one without) in units of the conductance quantum $G_0$, as a function of the tip position for the case of a metallic armchair nanoribbon (with $W=600\, a_0$),  for a Fermi energy that places the unperturbed conductance in the second plateau (point $P_2$ of Fig.~\ref{fig:TvsE}). 

The SGM scan of Fig.~\ref{fig:SGMscan_strip}  appears as translationally invariant in the longitudinal direction,  despite two effects that break this symmetry.  On the one hand,  the lattice-induced symmetry breaking in the $y$ direction produces conductance modulations that are imperceptible  within the chosen scale.  On the other hand,  cutting the tip-potential tail at the leads results in a finite-size effect that we minimize by simulating a strip which is more than seven times longer than the sector shown. 

While many features of quantum transport through graphene nanostructures can be inferred from the scans like that of Fig.~\ref{fig:SGMscan_strip},  the large number of physical parameters involved indicate that a systematic approach must be pursued. In the next section we develop a perturbative approach accounting for the SGM response of Dirac electrons in graphene to a noninvasive tip potential,  that will guide our discussion of the SGM results upon different conditions.

\section{Perturbative approach for noninvasive tips}
\label{sec:perturbative}

The effect of a noninvasive tip can be obtained from perturbation theory following the same lines as in the case of the 2DEG \cite{Jalabert2010,Gorini2013}. We present in this section the main steps of the approach, while technical details can be found in Appendixes \ref{sec:appendixA} and \ref{sec:appendixB}. 

The Dyson equation for the 
total Green function $\mathcal{G}_\mathrm{T}$ (including the effect of the tip) can be written in terms of the unperturbed Green function $\mathcal{G}$ and the tip-induced potential $U_\mathrm{T}$ as
\begin{align}
\label{eq:Dyson}
\mathcal{G}_\mathrm{T}(\br;\brp;\varepsilon)=&\;
\mathcal{G}(\br;\brp;\varepsilon)
\nonumber\\
&+
\int \dif \brpp \ 
\mathcal{G}(\br;\brpp;\varepsilon) \
\UT(\brpp) \
\mathcal{G}_\mathrm{T}(\brpp;\brp;\varepsilon) 
\,  .
\end{align}
Staying up to first order in $U_\mathrm{T}$ (i.e., within the Born approximation) and using the  Fisher-Lee relation \eqref{Fisher_Lee_graphene},  the correction to the transmission amplitude can be written as
\begin{widetext}
\begin{align}
\label{eq:deltat1}
\delta t_{ba}^{(1)} &=2\pi  {\rm i} (\hbar \vf)^2 \sum_{m,m^{\prime}=0}^{M+1} \varphi_{2,\varepsilon,b}^{(+)\dagger}(m,y) \, \sigma_y \left[ \sum_{m^{\prime\prime}}\int \dif y^{\prime\prime} \, \mathcal{G}(m,y;m^{\prime\prime},y^{\prime\prime};\varepsilon) \,
\UT(m^{\prime\prime}a_0,y^{\prime\prime}) \, \mathcal{G}(m^{\prime\prime},y^{\prime\prime};m^{\prime},y^{\prime},\varepsilon) \right] \sigma_y \, \varphi_{1,\varepsilon,a}^{(-)}(m^{\prime},y^{\prime})
\nonumber\\
&=2\pi  {\rm i} (\hbar \vf)^2 \sum_{m,m^{\prime}=0}^{M+1}
\sum_{\barl,\barl^{\prime}=1}^{2} 
\int_{-\infty}^{\infty} 
\frac{\dif \barep}{\varepsilon^{+}-\barep} 
\int_{-\infty}^{\infty}\frac{\dif \barep^{\prime}}{\varepsilon^{+}-\barep^{\prime}} 
\nonumber\\
&\quad\times \sum_{c,c^{\prime}}\varphi_{2,\varepsilon,b}^{(+)\dagger}(m,y) \ \sigma_y \ \Psi_{\barl,\barep,c}(m,y) \ 
\mathcal{U}_{c,c^{\prime}}^{\barl,\barl^{\prime}}(\barep,\barep^{\prime})
\Psi_{\barl^{\prime},\barep^{\prime},c^{\prime}}^{\dagger}(m^{\prime},y^{\prime}) \ \sigma_y \ 
\varphi_{1,\varepsilon,a}^{(-)}(m^{\prime},y^{\prime}) \, ,
\end{align}
where we  have skipped its explicit $\varepsilon$ dependence,  used the spectral decomposition \eqref{Green_function} of the unperturbed Green functions,  and defined the matrix element of the tip potential between scattering states
\begin{equation}
\label{eq:meotp}
\mathcal{U}_{c,c^{\prime}}^{\barl, \barl^{\prime}}(\barep,\barep^{\prime})=\sum_{m^{\prime\prime}=0}^{M+1} \int \dif y^{\prime\prime} \Psi_{\barl,\barep,c}^{\dagger}(m^{\prime\prime},y^{\prime\prime})\ \UT(m^{\prime\prime}a_0,y^{\prime\prime}) \ \Psi_{\barl^{\prime},\barep^{\prime},c^{\prime}}(m^{\prime\prime},y^{\prime\prime}) \,  .
\end{equation}
In Appendix \ref{sec:appendixC} we present explicit calculations for the case where the system is a GNR [see Eqs.~\eqref{eq:meotp4} and \eqref{eq:meotp7}].

Assuming a smooth dependence of the tip-potential matrix elements on $\barep$ and $ \barep^{\prime}$,  expressing the scattering states $\Psi$ in terms of the modes $\varphi^{(\mp)}$, and using the result \eqref{eq:energyintegrations},  we have
\begin{subequations}
\label{eq:energyintegrationsU}
\begin{align}
\label{eq:energyintegrationsU1}
& \sum_{m=0}^{M+1} \int_{-\infty}^{\infty} 
\frac{\dif \barep}{\varepsilon^{+}-\barep} \
\varphi_{2,\varepsilon,b}^{(+)\dagger}(m,y) \ \sigma_y \ 
\Psi_{\barl,\barep,c}(m,y) \
\mathcal{U}_{c,c^{\prime}}^{\barl, \barl^{\prime}}(\barep,\barep^{\prime}) 
= - \frac{\rm i}{\hbar \vf} 
\left( t_{bc}\delta_{\barl,1} + r^{\prime}_{bc}\delta_{\barl,2} \right) \mathcal{U}_{c,c^{\prime}}^{\barl, \barl^{\prime}}(\varepsilon,\barep^{\prime}) 
 \,  ,
\\
\label{eq:energyintegrationsU2}
& \sum_{m'=0}^{M+1} 
\int_{-\infty}^{\infty}\frac{\dif \barep^{\prime}}{\varepsilon^{+}-\barep^{\prime}} \
\Psi_{\barl^{\prime},\barep^{\prime},c^{\prime}}^{\dagger}(m^{\prime},y^{\prime}) \ \sigma_y \ 
\varphi_{1,\varepsilon,a}^{(-)}(m',y') \
\mathcal{U}_{c,c^{\prime}}^{\barl, \barl^{\prime}}(\barep,\barep^{\prime})  =
- \frac{\rm i}{\hbar \vf} \delta_{c',a} \ \delta_{\barl^{\prime},1} \
\mathcal{U}_{c,c^{\prime}}^{\barl, \barl^{\prime}}(\barep,\varepsilon) 
 \,  .
\end{align}
\end{subequations}

\end{widetext}
Therefore,
\begin{equation}
\label{eq:deltat1final}
\delta t_{ba}^{(1)} =- 2\pi {\mathrm i} \sum_{c}\left[t_{bc} \ \mathcal{U}_{c,a}^{1,1}(\varepsilon,\varepsilon)+r_{bc}^{\prime} \ \mathcal{U}_{c,a}^{2,1}(\varepsilon,\varepsilon) \right] \,  ,
\end{equation}
and we obtain the first-order correction to the conductance 
\begin{equation}
\label{eq:foc}
\frac{\delta G^{(1)}}{G_0} = 2 \sum_{a,b} {\rm Re} \left\{t_{ba}^{*} \ \delta t_{ba}^{(1)} \right\} = 4\pi \ {\rm Im} \left\{{\rm Tr} \left[t^{\dagger} r^{\prime} \mathcal{U}^{2,1} \right] \right\} 
 \,  ,
\end{equation}
which has the same form as in the case of the 2DEG \cite{Jalabert2010}.  Expressing the conductance correction as a trace,  analogously to the Landauer formula \eqref{eq:conductance}, makes it manifestly independent of the basis chosen for the lead states.  The basis of the transmission eigenmodes  is quite useful since it allows to write Eq.~\eqref{eq:foc} as a single sum over the propagating eigenmodes \cite{Gorini2013}
\begin{equation}
\label{eq:focmd}
\frac{\delta G^{(1)}}{G_0} = 4\pi \ \sum_{n=1}^{N} \ {\cal T}_n  \ {\cal R}_n \ {\rm Im} \left\{\mathcal{U}_{n,n}^{2,1}  \right\} 
 \,   ,
\end{equation}
where ${\cal T}_n$ (${\cal R}_n$) is the transmission (reflection) eigenvalue and $\mathcal{U}_{n,n}^{2,1}$ is the diagonal matrix element of the tip potential in the basis of the transmission eigenmodes.  The case of a strip is particularly simple since the lead modes \eqref{eq:modes} are the transmission eigenmodes.

It follows directly from Eq.~\eqref{eq:foc} that there is no first-order correction for the perfectly transmitting modes.  In these cases, which include that of the GNRs,  we need to go beyond the first-order approximation in order to address the SGM response. The second-order SGM conductance correction can be written as
\begin{equation}
\label{eq:deltaG2}
\delta G^{(2)} = \delta G^{(2)\alpha} + \delta G^{(2)\beta} \, ,
\end{equation} 
with
\begin{subequations}
\label{eq:deltaG2d}
\begin{align}
\label{eq:deltaG2alpha}
\frac{\delta G^{(2)\alpha}}{G_0} &=  2 \ {\rm Re} \left\{{\rm Tr} \left[t^{\dagger} \delta t^{(2)}  \right] \right\} 
 \,  ,
\\
\label{eq:deltaG2beta}
\frac{\delta G^{(2)\beta}}{G_0} &=   {\rm Tr} \left[ {\delta t^{(1)}}^{\dagger} \delta t^{(1)}  \right]
 \,  .
\end{align}
\end{subequations}
Going up to second order in Eq.~\eqref{eq:Dyson} we have
\begin{widetext}
\begin{align}
\label{eq:deltat2}
\delta t_{ba}^{(2)}=&\; 2\pi{\rm i} (\hbar  \vf)^{2}
\sum_{m,m^{\prime}=0}^{M+1} \
\sum_{\barl,\barl^{\prime},\barl^{\prime\prime}=1}^{2} 
 \int_{-\infty}^{\infty} 
\frac{\dif \barep}{\varepsilon^{+}-\barep} 
\int_{-\infty}^{\infty}\frac{\dif \barep^{\prime}}{\varepsilon^{+}-\barep^{\prime}}
\int_{-\infty}^{\infty}\frac{\dif \barep^{\prime\prime}}{\varepsilon^{+}-\barep^{\prime\prime}}  \nonumber \\
&\times\sum_{c,c^{\prime},c^{\prime\prime}} \varphi_{2,\varepsilon,b}^{(+)\dagger}(m,y) \ \sigma_y \ 
\Psi_{\barl,\barep,c}(m,y) \
\mathcal{U}_{c,c^{\prime}}^{\barl,\barl^{\prime}}(\barep,\barep^{\prime}) \
\mathcal{U}_{c^{\prime},c^{\prime\prime}}^{\barl^{\prime},\barl^{\prime\prime}}(\barep^{\prime},\barep^{\prime\prime}) \
\Psi_{\barl^{\prime\prime},\barep^{\prime\prime},c^{\prime\prime}}(m^{\prime},y^{\prime}) \ \sigma_y \ \varphi_{1,\varepsilon,a}^{(-)}(m^{\prime},y^{\prime}) \, .
\end{align}
The result of the $m$ sum and $\barep$ integral of  the above equation follows from Eq.~\eqref{eq:energyintegrationsU1},  and similarly,  the $m'$ sum and $\barep^{\prime\prime}$ integration from 
Eq.~\eqref{eq:energyintegrationsU2},  leading to
\begin{equation}
\label{eq:deltat2bis}
\begin{aligned}
\delta t_{ba}^{(2)}&= - 2\pi {\rm i} \sum_{\barl^{\prime}}
\int_{-\infty}^{\infty}\frac{\dif \barep^{\prime}}{\varepsilon^{+}-\barep^{\prime}}
\sum_{c,c^{\prime}}
\left[ t_{bc} \
\mathcal{U}_{c,c^{\prime}}^{1,\barl^{\prime}}(\varepsilon,\barep^{\prime}) \
\mathcal{U}_{c^{\prime},a}^{\barl^{\prime},1}(\barep^{\prime},\varepsilon) + r_{bc}^{\prime} \
\mathcal{U}_{c,c^{\prime}}^{2,\barl^{\prime}}(\varepsilon,\barep^{\prime}) \
\mathcal{U}_{c^{\prime},a}^{\barl^{\prime},1}(\barep^{\prime},\varepsilon)
\right].\\
\end{aligned}
\end{equation}
We thus have 
\begin{equation}
\begin{aligned}
\frac{\delta G^{(2)\alpha}}{G_0}&= -4\pi^2 
\sum_{\barl^{\prime}=1}^{2}
\mathrm{Tr}
\left[ t^{\dagger} \ t \ \mathcal{U}^{1\barl^{\prime}} \
\mathcal{U}^{\barl^{\prime}1} \right] + 4\pi \sum_{\barl^{\prime}=1}^{2}
\mathrm{Im}
\int_{-\infty}^{\infty}\frac{\dif \barep^{\prime}}{\varepsilon^{+}-\barep^{\prime}}
\mathrm{Tr}\left[t^{\dagger} \ r^{\prime} \
\mathcal{U}_{c,c^{\prime}}^{2,\barl^{\prime}}(\varepsilon,\barep^{\prime}) \
\mathcal{U}_{c^{\prime},a}^{\barl^{\prime},1}(\barep^{\prime},\varepsilon)
\right] \,  ,
\end{aligned}
\end{equation}
where we have used that $\mathrm{Tr}
\left[ t^{\dagger} \ t \ \mathcal{U}^{1,\barl^{\prime}} \
\mathcal{U}^{\barl^{\prime},1} \right]$ is real.  The correction $\delta G^{(2)\beta}$ can be readily obtained from Eqs.~\eqref{eq:deltaG2beta} and \eqref{eq:deltat1final},  leading to 
\begin{equation}
\frac{\delta G^{(2)\beta}}{G_0} = 4\pi \mathrm{Tr} \left[ t^{\dagger} \  t\ \mathcal{U}^{1,1} \ \mathcal{U}^{1,1} + r^{\prime\dagger} \ r^{\prime} \ \mathcal{U}^{2,1} \ \mathcal{U}^{1,2} + 2 \mathrm{Re}\left\{ r^{\prime\dagger} \ t \ \mathcal{U}^{1,1} \ \mathcal{U}^{1,2} \right\} \right] \, .
\end{equation}

From $\delta G^{(2)\alpha}$ and $\delta G^{(2)\beta}$ we obtain the second-order conductance correction 
\begin{align}
\label{eq:soc}
\frac{\delta G^{(2)}}{G_0} =& -  4\pi^2
\mathrm{Tr}\left[
t^{\dagger} \ t \ \mathcal{U}^{1,2} \ \mathcal{U}^{2,1} -
r^{\prime\dagger} \ r^{\prime} \ \mathcal{U}^{2,1} \ \mathcal{U}^{1,2}-
2\mathrm{Re}\left\{ r^{\prime\dagger} \ t \ \mathcal{U}^{1,1} \ \mathcal{U}^{1,2} \right\} \right] 
\nonumber\\
&+ 4\pi \sum_{\barl^{\prime}=1}^{2}
\mathrm{Im}
\int_{-\infty}^{\infty}\frac{\dif \barep^{\prime}}{\varepsilon^{+}-\barep^{\prime}} \
\mathrm{Tr}\left[t^{\dagger} \ r^{\prime} \
\mathcal{U}^{2,\barl^{\prime}}(\varepsilon,\barep^{\prime}) \
\mathcal{U}^{\barl^{\prime},1}(\barep^{\prime},\varepsilon)
\right]
\, .
\end{align}
\end{widetext}

As for the first-order correction,  the representation of transmission eigenmodes considerably simplifies the above expression,  since the traces can be calculated as single sums.  In the case where all the propagating eigenmodes are perfectly transmitting,  as in GNRs,  Eq.~\eqref{eq:soc} reduces to 
\begin{equation}
\label{eq:socpt}
\frac{\delta G^{(2)}}{G_0} = -4\pi^2  \ \mathrm{Tr}
\left[\mathcal{U}^{1,2} \ \mathcal{U}^{2,1} \right] \, .
\end{equation}
This second-order correction to the conductance in the case of perfect transmission is trivially negative (or null),  respecting the constraint that the presence of the tip cannot open additional conductance channels in the leads,  but only reduce the transmission of the existing channels. 

Equations \eqref{eq:foc} and \eqref{eq:socpt} are the basis of our analysis of the SGM patterns obtained in GNRs and graphene QPCs for noninvasive tips,  and will guide our discussion of the numerically obtained SGM results in different conditions.

\section{\label{sec:correction_ribb}SGM correction in a metallic armchair graphene nanoribbon}
 
We first tackle the geometry of a nanoribbon,  which is particularly simple since the perfect transmission of the propagating modes results in a vanishing first-order conductance correction, i.e.,  $\delta G^{(1)}=0$ according to Eq.~\eqref{eq:foc}.  Therefore,  the second-order term $\delta G^{(2)}$,  given by Eq.~\eqref{eq:socpt},  is the leading-order correction for a noninvasive probe.  Moreover,  in the case of armchair GNRs, the tip-potential matrix elements can be calculated under some approximations (see Appendix \ref{sec:appendixC}).  For the case of $d \ll W$ and distances from the tip to the boundaries larger than $d$,  the matrix element \eqref{eq:meotp4} is independent of the tip position,  resulting in 
\begin{equation}
\label{eq:socpt2}
\frac{\delta G^{(2)}}{G_0} = 
- \left(\frac{2 \pi u_{\rm T} d^2}{\hbar \vf W}\right)^2
\sum_{a} \left(\frac{q_a}{k_a}\right)^2
K_0^2(2k_a d)
 \,  ,
\end{equation}
where the sum is over the propagating modes $a$,  and $K_0$ stands for the zeroth-order modified Bessel function of the second kind.

\begin{figure}[tb]
 \includegraphics[width=\linewidth]{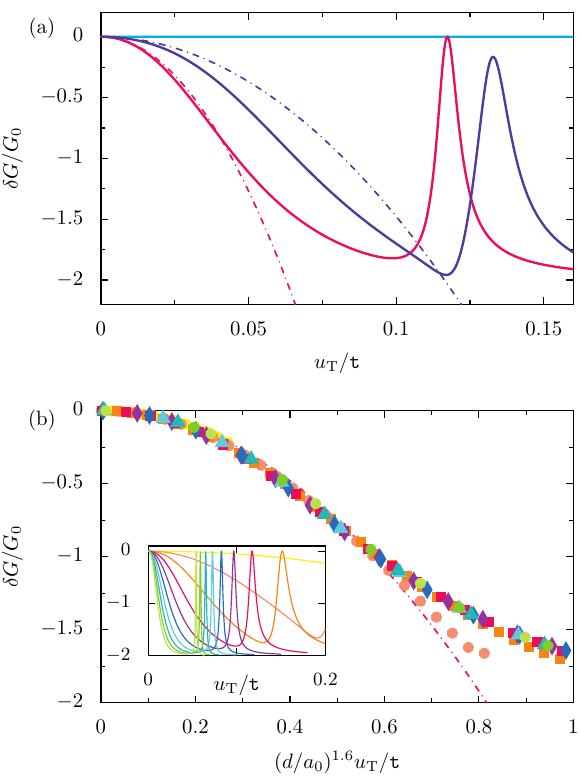}
\caption{(a) SGM conductance correction for a metallic armchair GNR of width $W_\mathrm{L}=599\, a_0$ when the tip is placed on the axis ($x=W/2$) of the strip, for three cases where the unperturbed conductance is in the first (blue solid line), second (red solid line), and third (purple solid line) plateau (points $P_1$, $P_2$, and $P_3$ of Fig.~\ref{fig:TvsE},  respectively) as a function of the strength $u_{\rm T}$ of the tip-generated potential  (scaled by the hopping constant $\mathtt{t}$). The extent of the tip-induced potential is given by the value $d=20\, a_0$. The red and purple dash-dotted lines represent the analytical predictions of Eq.~\eqref{eq:socpt2} for the initial quadratic dependence on $u_{\rm T}$ for the cases of $P_2$ and $P_3$, respectively. (b) Data collapse of the initial correction for the point $P_2$ of the second plateau obtained for various tip sizes $5 \leqslant d/a_0 \leqslant 50$ and strengths $0 \leqslant u_{\rm T}/\mathtt{t} \leqslant 0.2$, once the scaling $u_{\rm T} d^{1.6}$ is implemented. The unscaled data are shown in the inset as a function of $u_\mathrm{T}$ (for values including the initial decrease and the first revival peak). The colors correspond to those of the data points in the main plot. Curves with increasing values of $d$ show stronger and stronger conductance corrections.} 
 \label{fig:SGMfV}
\end{figure}

In Fig.~\ref{fig:SGMfV}(a), we present the numerically obtained SGM correction as a function of the tip strength $u_{\rm T}$ (scaled by the hopping constant $\mathtt{t}$) when the tip of fixed size $d = 20\, a_0$ is placed on the axis of the strip (at $x=W/2$ with the coordinate system of Fig.~\ref{fig:lattice}) for three cases where the unperturbed conductance is in the first (blue solid line), second (red solid line), and third (purple solid line) plateau (points $P_1$,  $P_2$,  and $P_3$ of Fig.~\ref{fig:TvsE},  respectively). 

The first remarkable feature is that when only the zero-transverse-energy mode is occupied, the SGM correction vanishes on the scale of the other curves (typically, $\delta G/G_0\sim-10^{-8}$).  This singularity, already noticed in Ref.~\cite{Mrenca_2015,*Mre_ca_Kolasi_ska_2022},  is consistent with Eq.~\eqref{eq:socpt2}, which dictates that the second-order conductance correction trivially vanishes in this case.  Moreover, as we stress in Appendix \ref{sec:appendixC},  it can be shown that $\mathcal{U}_{a^{*},a^{*}}^{2,1}(\varepsilon,\varepsilon)=0$, without introducing all the approximations that lead to Eq.~\eqref{eq:socpt2},  and independently of the characteristics of the tip potential ($u_{\rm T}$ and $d$),  provided it is long ranged.  The result is indeed valid for an arbitrary long-range disorder potential. This leads to the concept of nearly perfect single-channel conduction for armchair GNRs
\cite{Yamamoto2009}, as the higher-order terms in the perturbation expansion cannot in principle be ruled out, since other matrix elements do not vanish [see Eq.~\eqref{eq:meotp7}].  Invoking the time-symmetry breaking operated by the boundary conditions in the case of metallic armchair GNRs, Ref.~\cite{Wurm2012} proposed that the zero-transverse-energy mode in armchair GNRs is a perfectly conducting channel.

The distinction between a perfectly conducting channel and a nearly perfect conducting one is delicate from the numerical point of view 
\cite{Wurm2012}.  Indeed, the small values of $\delta G$ in Fig.~\ref{fig:SGMfV}(a) are affected by the finite-size effects due to the long-range tip potential extending up to the leads in our simulations with a finite extent along the $y$ direction, as well as by the fact that the potential of the numerical calculations has a finite range and therefore mixes the graphene valleys.  However, from the theoretical point of view, the perfect conduction of the zero-transverse-energy channel can be addressed by going to higher order in the perturbation expansion.  

For the zero-transverse-energy mode, the first corrections to the transmission amplitude are easily obtained from Eqs.~\eqref{eq:deltat1final},  \eqref{eq:deltat2bis},  and \eqref{eq:energyintegrationsZTEM}, resulting in  $\delta t_{a^{*}a^{*}}^{(1)}=-2 \pi {\rm i} \ \mathcal{U}_{a^{*},a^{*}}^{1,1}(\varepsilon,\varepsilon) $ and $\delta t_{a^{*}a^{*}}^{(2)}=-2 \ [\pi \ \mathcal{U}_{a^{*},a^{*}}^{1,1}(\varepsilon,\varepsilon)]^2 $.  It is easy to see that the $n$th-order correction to the transmission amplitude $\delta t_{ba}^{(n)}$ has the same structure as Eq.~\eqref{eq:deltat2bis},  with $n-1$ intermediate energy integrals and sums over the lead index, and having products of $n$ matrix elements in each term.  For the zero-transverse-energy mode important simplifications appear from the condition $\mathcal{U}_{a^{*},a^{*}}^{2,1}(\varepsilon,\varepsilon)=0$.  From the general result \eqref{eq:energyintegrationsZTEMn},  we have
\begin{equation}
\label{eq:deltatnZEM}
\delta t_{a^{*}a^{*}}^{(n)}=2 \ \left[- {\rm i} \pi \ \mathcal{U}_{a^{*},a^{*}}^{1,1}(\varepsilon,\varepsilon)\right]^n \,  ,
\end{equation}
which allows us to calculate corrections $\delta G^{(n)}$ of arbitrary order $n$.  For odd $n$ we have
\begin{equation}
\frac{\delta G^{(n)}}{G_0} =  
2 \ {\rm Re}  \left[\delta t_{a^{*}a^{*}}^{(n)} + \sum_{j=1}^{(n-1)/2}\left[\delta t_{a^{*}a^{*}}^{(j)}\right]^{*} \delta t_{a^{*}a^{*}}^{(n-j)} \right] = 0 \,  ,
\end{equation}
since $\delta t^{(n)}$,  as well as each term of the sum,  are all pure imaginary.  For even $n$ we have a decomposition analogous to Eq.~\eqref{eq:deltaG2} with
\begin{subequations}
\label{eq:deltaG4d}
\begin{align}
\label{eq:deltaG4alpha}
\frac{\delta G^{(n)\alpha}}{G_0} &= 
 2 \ {\rm Re}  \left[\delta t^{(n)}_{a^{*}a^{*}} + \sum_{j=1}^{n/2-1}\left[\delta t^{(j)}_{a^{*}a^{*}}\right]^{*} \delta t^{(n-j)}_{a^{*}a^{*}} \right]
 \,  ,
\\
\label{eq:deltaG4beta}
\frac{\delta G^{(n)\beta}}{G_0} &=    \left| {\delta t^{(n/2)}_{a^{*}a^{*}}} \right|^2
 \,  .
\end{align}
\end{subequations}
Using Eq.~\eqref{eq:deltatnZEM} we have
\begin{align}
\frac{\delta G^{(n)\alpha}}{G_0} &=  4 \ 
\left[ {\rm i} \pi \ \mathcal{U}_{a^{*},a^{*}}^{1,1}(\varepsilon,\varepsilon)\right]^n
\left[1+ 2 \sum_{j=1}^{n/2-1}(-1)^{j} \right]
 \,  \nonumber
\\
 &=   - 4 \left[ \pi \ \mathcal{U}_{a^{*},a^{*}}^{1,1}(\varepsilon,\varepsilon)\right]^n  = - \frac{\delta G^{(n)\beta}}{G_0} 
 \,  .
\end{align}

We therefore have $\delta G=0$ to all orders in perturbation theory, implying that the zero-transverse-energy mode is a perfectly conducting channel.  We stress that this result is not restricted to the particular form \eqref{eq:meotp7} of a matrix element corresponding to a Lorentzian-shaped tip potential,  but it applies to any long-range potential,  including the disordered case treated in Ref.~\cite{Wurm2012}.

The SGM response for the points $P_2$ and $P_3$, on the second and third plateaus, depicted by red and purple solid lines in Fig.~\ref{fig:SGMfV}(a), respectively,  exhibits an initial quadratic dependence as a function of the potential strength $u_{\rm T}$. The corrections are always negative, in agreement with the expectation that the dominant SGM correction at weak tip strength is of second order, and the initial strength dependencies are well described by the perturbative prediction of Eq.~\eqref{eq:socpt2} (red and purple dash-dotted lines, respectively).  It is important to remark that the perturbative regime extends over a relatively large $u_{\rm T}$ interval and describes rather precisely conductance corrections up to $|\delta G|\lesssim G_0$.  Moreover, the SGM scan of Fig.~\ref{fig:SGMscan_strip} confirms the prediction of Eq.~\eqref{eq:socpt2} of an approximate independence with respect to the tip position,  provided that the walls are not approached.

In Fig.~\ref{fig:SGMfV}(b) we present the SGM correction for different potential widths $d$ (different symbols) and strengths $u_{\rm T}$ in a large range ($5 \leqslant d/a_0 \leqslant 50$,  $0 \leqslant u_{\rm T}/\mathtt{t} \leqslant 0.2$), which demonstrates a robust data collapsing. The noninteger power law for the scaling in the variable $d$ is consistent with the logarithmic dependence of the function $K_0$ in Eq.~\eqref{eq:socpt2} for small values of the argument.

\begin{figure}
 \includegraphics[width=\linewidth]{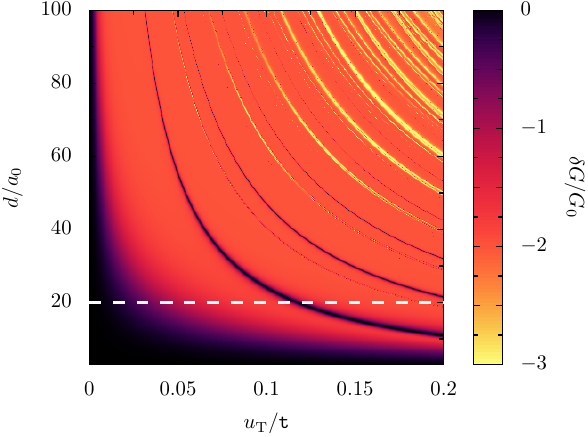}
\caption{SGM conductance correction (in units of the conductance quantum $G_0£$) for a metallic armchair GNR, when the tip is placed on the axis of the strip ($x=W/2$) and the unperturbed conductance is in the second plateau (point $P_2$ of Fig.~\ref{fig:TvsE}),  as a function of the strength ($u_{\rm T}$) and extent ($d$) of the tip-generated potential. On the dashed white horizontal line corresponding to $d=20\, a_0$, the data of Fig.~\ref{fig:SGMfV}(a) for the second plateau (red solid line) are reproduced.} 
 \label{fig:SGMfVandD}
\end{figure}

Another remarkable feature of Fig.~\ref{fig:SGMfV}(a) is the revival of the GNR conductance for large values of $u_\mathrm{T}$, outside the perturbative regime.  These peaks attain the unitary limit of no conductance correction in the case of the second plateau (red solid line), and can be understood as resonances through states electrostatically confined below the tip in the $p$ region defined for sufficiently large values of $u_\mathrm{T}$, like the graphene quantum-dot states studied in Refs.~\cite{Downing2011,Brouwer2014}. In Fig.~\ref{fig:SGMfVandD} the SGM conductance correction is presented in color scale as a function of tip strength and width, highlighting the resonances (dark colors) and antiresonances (light colors) that appear as functions of $u_{\rm T}$ and $d$ when the unperturbed conductance is in the second plateau (point $P_2$ of Fig.~\ref{fig:TvsE}). These resonance lines indicate the relation between $u_\mathrm{T}$ and $d$, under which the confined states under the tip remain aligned with the Fermi energy.  
The cut shown by the dashed white horizontal line corresponding to $d=20\, a_0$ indicates the parameters for which the data have been presented in Fig.~\ref{fig:SGMfV}(a) for the second plateau (red solid line), on a slightly larger $u_{\rm T}$ interval.

The resonances occur at the same tip strengths when the tip position is moved away from the nanoribbon axis.  When the edges of the ribbon are approached, the revival occurs at lower tip strength as it can be expected from the behavior of confined states in a truncated potential.

\section{\label{sec:correction_qpc}SGM correction in a graphene nanoconstriction}

\begin{figure}[tb]
\includegraphics[width=\linewidth]{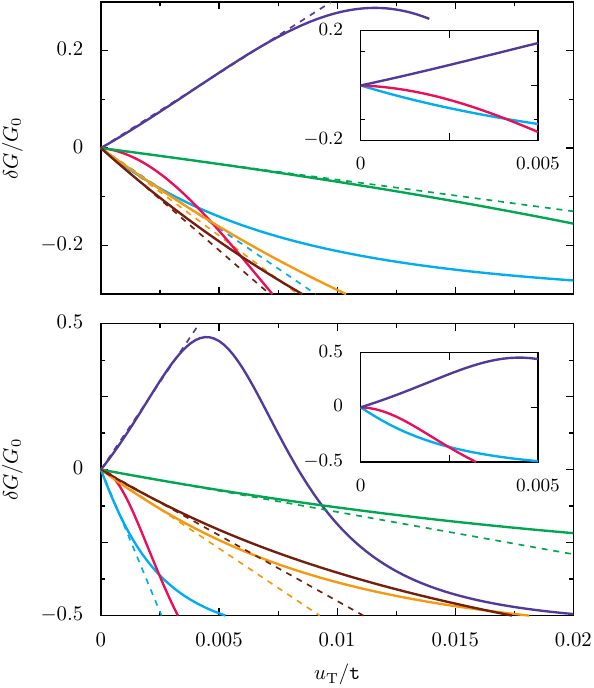}
\caption{Conductance change in a GNC defined by Eq.~\eqref{eq:QPCdef},  for $W_\mathrm{C}=399\, a_0$ and $C=\unit[25]{nm}$ (upper panel),  together with the case of $W_\mathrm{C}=199\, a_0$ and $C=\unit[25]{nm}$ (lower panel),  as a function of the tip strength,  for a tip of width $d= 20 \, a_0$ placed at the center of the constriction.  The different solid lines correspond to the unperturbed conditions characterized by the points $Q_i$ ($S_i$), of the same color,  in the upper (lower) panel of Fig.~\ref{fig:TvsEmd}.  The dashed lines represent linear fits to the behavior at low tip strength.  Insets: Detail of the small $u_\mathrm{T}$ region for the conditions defined by the points $Q_i$ (upper panel) and $S_i$ (lower panel),  with $i=1,2,3$ characterizing a peak of the unperturbed conductance.}
\label{fig:SGMQPC}
\end{figure}
In the case of a GNC defined in a nanoribbon we do not have well-defined conductance plateaus (see colored solid lines in Fig.\ \ref{fig:TvsE}),  and therefore,  we expect the lowest nonvanishing SGM conductance correction to be the first-order term of Eq.~\eqref{eq:foc}, yielding a linear dependence on tip strength in the noninvasive regime. The numerical results of Fig.\ \ref{fig:SGMQPC}, showing the dependence of the SGM correction as a function of the tip strength for two GNCs shown in Fig.~\ref{fig:TvsE} ($W_\mathrm{C}=399\, a_0$ and $C=\unit[25]{nm}$ for the upper panel and $W_\mathrm{C}=199\, a_0$ and $C=\unit[25]{nm}$  for the  lower panel) and a tip placed at their center confirm our expectation for all the operating conditions defined in Fig.~\ref{fig:TvsEmd}, except for the points $Q_2$ and $S_2$ characterizing a maximum of the unperturbed conductance.  

Unlike the negative second-order correction that dominates in the conductance plateaus, the first-order correction, indicated by fits of linear $u_\mathrm{T}$ dependence (dashed lines with the corresponding color), can be positive or negative.  The matrix elements \eqref{eq:meotp} that, together with the unperturbed transmission and reflection amplitudes, determine the value of the conductance correction are more difficult to evaluate than for the case of GNRs,  since the scattering wave functions are in general not known (except for particular cases like those of abrupt junctions \cite{Wurm2009}).  It can be observed in Fig.~\ref{fig:SGMQPC} that the range of linear behavior is quite reduced,  as higher-order terms become relevant already at moderate tip strengths of $u_\mathrm{T}\approx 0.01 \, \mathtt{t}$. 

The points $Q_2$ and $S_2$ in Fig.~\ref{fig:TvsEmd} (red lines in Fig.~\ref{fig:SGMQPC}) correspond to unperturbed unitary transmissions set by the fictitious GNRs of width $W_\mathrm{C}$.  The effect of the tip can thus only reduce the conductance,  and therefore  the second-order correction dominates,  resulting in the quadratic dependence for very small values of $u_\mathrm{T}$ observed in Fig.\ \ref{fig:SGMQPC}.  Points like $Q_5$ and $S_5$ (green lines), corresponding to a plateaulike condition in Fig.\ \ref{fig:TvsEmd},  exhibit a very small slope,  indicating that the linear correction \eqref{eq:foc} is weakened when approaching the regime of conductance quantization.  The overall conductance scale in Fig.~\ref{fig:SGMQPC} of the upper panel is smaller than that of the lower panel,  since the former presents the case of a smoother junction than the latter and the well-defined quantized conductance plateaus of the smooth junction weaken the contribution arising from the first-order correction \eqref{eq:focmd}.

\begin{figure}[tb]
  \includegraphics[width=\linewidth]{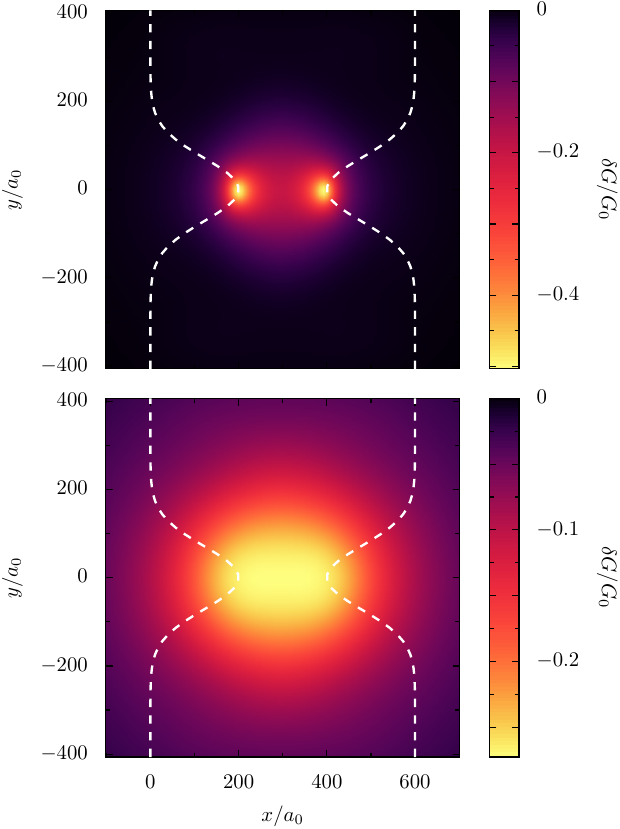} 
   \caption{SGM scan for a GNC defined in a nanoribbon by Eq.~\eqref{eq:QPCdef},  with $W=600\, a_0$, $W_\mathrm{C}=199\, a_0$, and $C=\unit[25]{nm}$ (white dashed lines),  for the unperturbed condition corresponding to the point $S_4$ in the lower panel of Fig.~\ref{fig:TvsEmd}. The tip potential strength is $u_{\rm T}=0.002\, \mathtt{t}$, while the tip extent is $d=20\, a_0$ ($d=100\, a_0$) for the upper (lower) panel.}
  \label{fig:SGMscan_QPC} 
\end{figure}

Figure \ref{fig:SGMscan_QPC} shows the scan of the conductance correction in a GNC,  as a function of tip position,  with an unperturbed condition corresponding to the point $S_4$ (orange) of Fig.~\ref{fig:TvsEmd}(b) for the case of a weak tip in the noninvasive regime ($u_\mathrm{T} = 0.002 \, \mathtt{t}$) and two values of the tip-potential extent: $d=20\, a_0$ (upper panel) and $d=100\, a_0$ (lower panel).  In both cases, the strongest SGM response appears when the tip is close to the GNC, in agreement with previous experimental \cite{Neubeck_2012} and theoretical work \cite{Mrenca_2015,*Mre_ca_Kolasi_ska_2022,Kolasi2017_b}.  Such strong response overwhelms, on the scale of the figures,  the modulations seen in Fig.~\ref{fig:SGMscan_strip} arising from the GNR enclosure.  On the one hand,  the tip strength is much smaller than that used in the example of SGM in a GNR.  On the other hand,  the point $S_4$ corresponds to the third plateau of the limiting GNR,  described by the point $P_3$ in Fig.~\ref{fig:TvsE},  where the SGM response is considerably weaker than in the second plateau (point $P_2$).  The smaller tip ($d=20\, a_0$) results in a weaker signal than the large tip ($d=100\, a_0$) since the potential \eqref{eq:tip_potential} becomes less effective as $d$ decreases [see Eq.~\eqref{eq:meotp4}].  We notice that the SGM conductance correction is not only negative at the center of the constriction [orange line in Fig.~\ref{fig:SGMQPC}(b)],  but everywhere in the region scanned in Fig.~\ref{fig:SGMscan_QPC}.

\begin{figure}[tb]
  \includegraphics[width=\linewidth]{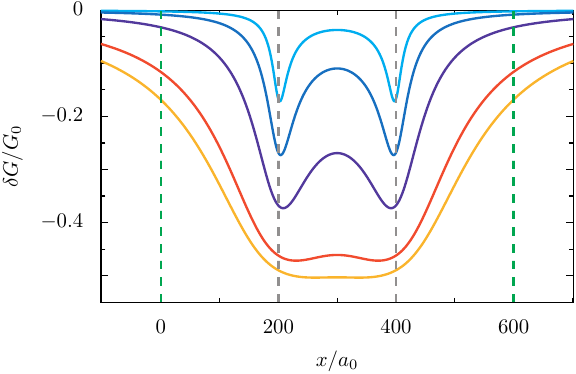} 
   \caption{SGM scan along the transverse central line ($y=0$) for the GNC of Fig.~\ref{fig:SGMscan_QPC} with different tip extents  ($d/a_0=10$, $20$, $40$, $80$, $100$ from top to bottom).  The dashed grey vertical lines indicate the position of the points where the constriction is the narrowest,  and the dashed green vertical lines stand for the limits of the nanoribbon on which the GNC is defined.}
  \label{fig:SGMQPCcut} 
\end{figure}

An important difference between the two panels of Fig.~\ref{fig:SGMscan_QPC} is that the SGM response corresponding to the small tip presents a spatial feature with two local minima close to the edges in the narrowest part of the GNC.  Figure \ref{fig:SGMQPCcut} shows the conductance change when the tip is displaced along the $y=0$ line for different widths ($d/a_0=10, 20,\dots , 100$),  illustrating how a progressively broader tip blurs the localized features at the edges.  The spatial features of Fig.~\ref{fig:SGMscan_QPC} are rather generic,  as they appear for most of the unperturbed conditions,  but other behavior can be also observed;  i.e.,  for the point $S_1$ (not shown) no concentration of the SGM signal is obtained.

\begin{figure}[tb]
  \includegraphics[width=\linewidth]{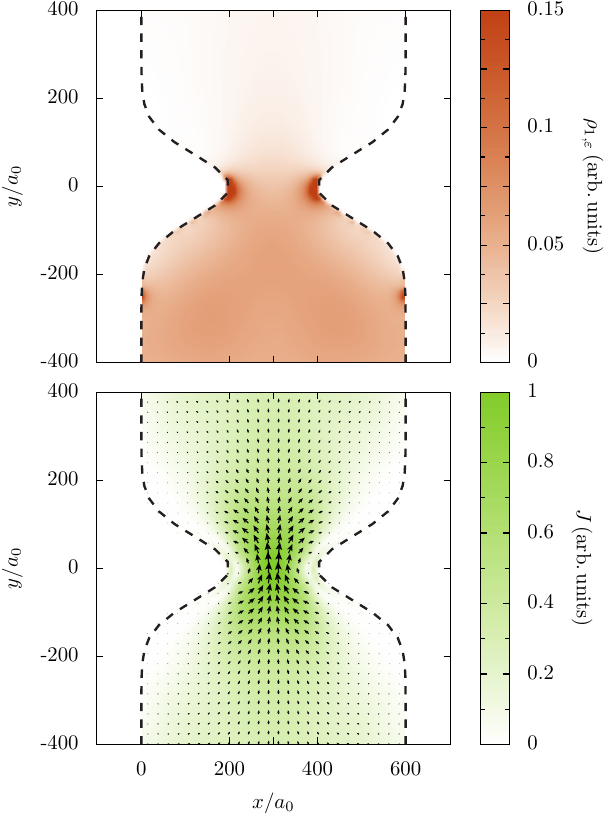} 
   \caption{Top: Partial local density of states impinging from the lower lead \eqref{eq:pldos}, calculated for the GNC of Fig.~\ref{fig:SGMscan_QPC} (in arbitrary units). Bottom: Corresponding current density.}
  \label{fig:PLDOS} 
\end{figure}

In semiconductor QPCs, it has been shown \cite{Ly2017} that, in certain cases,  a weakly invasive SGM response can be connected with the unperturbed partial local density of states (PLDOS), defined as
\begin{equation}\label{eq:pldos}
\rho_{1,\varepsilon}(\br)=2\pi \sum_{a=1}^{N}
|\Psi_{1,\varepsilon,a}(\br)|^2
\end{equation}
for electrons impinging into the scatterer from lead $1$ with an energy $\varepsilon$, and scattering states given by Eq.~\eqref{eq:scatteringstates1}.  For local tips, there is a linear relationship between the first-order conductance change and the PLDOS in the case of a single open channel,  while when the QPC is tuned on a conductance plateau,  the second-order conductance correction verifies 
$\delta G^{(2)}(\br_\mathrm{T})/G_0 = - \rho^2_{1,E_\mathrm{F}}(\br_\mathrm{T})$,  when the position $\br_\mathrm{T}$ of the tip is in the lead where electrons are transmitted. The departures from the above relationship for imperfect transmission were shown to be small, provided the extent of the tip remains small \cite{Ly2017}.  In the upper panel of Fig.~\ref{fig:PLDOS},  we explore the previous connection and plot the PLDOS of the GNC considered in Fig.~\ref{fig:SGMscan_QPC}, obtaining a large enhancement of the PLDOS at the edges in the narrowest part.  

An imaging technique using nitrogen-vacancy-center magnetometry has revealed that the current density is concentrated on the edges of the narrowest part of a micrometer-sized constriction operating in the Ohmic regime \cite{Jenkins2022}. When lowering the temperature below $\unit[200]{K}$, the Ohmic regime gives up its place to the ballistic regime and the current density becomes homogeneous along the centerline of the constriction. Our results for the current density (lower panel of Fig.~\ref{fig:PLDOS}) agree with the ballistic ones of Ref.~\cite{Jenkins2022}. Thus, the strong features appearing in the PLDOS and put in evidence by the SGM response to a relatively small tip do not seem to have an effect on the current density.

\section{Zigzag edges}
\label{sec:zigzag}

We have so far mainly discussed the case of armchair edges,  where the eigenenergies and eigenfunctions have the simple form presented in Appendix \ref{sec:appendixA},  and analytical expressions could be obtained in the perturbative regime.  The case of zigzag edges is more involved since in a GNR the possible transverse quantum numbers depend on the longitudinal momentum.  Another difference with respect to the armchair case is the existence in zigzag GNRs of flat bands with small values of $|\varepsilon|$,  corresponding to chiral eigenstates localized at the edges for each sublattice.   

\begin{figure}[tb]
  \includegraphics[width=\linewidth]{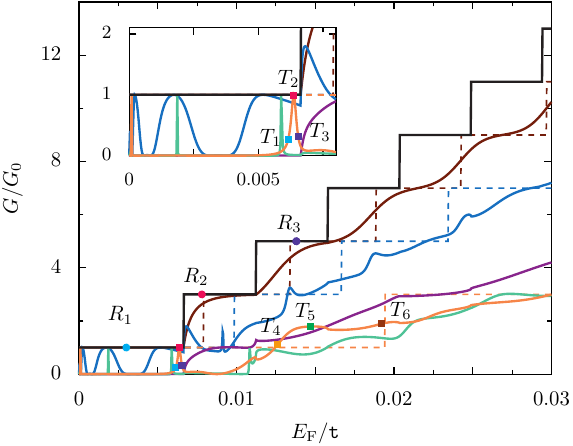} 
      \caption{Conductance as a function of the Fermi energy for different zigzag GNRs and GNCs. The thick solid black line corresponds to a GNR with a width $W_\mathrm{L}=599\, a_0$. The thick solid colored lines represent the conductance of GNCs of different shapes: $W_\mathrm{C}=499\, a_0$ and $C=\unit[25]{nm}$ (brown), $W_\mathrm{C}=399\, a_0$ and $C=\unit[25]{nm}$ (dark blue), $W_\mathrm{C}=199\, a_0$ and $C=\unit[40]{nm}$ (green),  $W_\mathrm{C}=199\, a_0$  and $C=\unit[25]{nm}$ (orange), $W_\mathrm{C}=199\, a_0$ and $C=\unit[10]{nm}$ (violet). The thin dashed lines stand for the conductance of GNRs with the width $W_\mathrm{C}$ of the corresponding GNC (according to the color convention). Inset: Detail of the low-energy sector of the main figure. The different points $R_i$ ($T_i$) define the parameters chosen to perform the SGM analysis of zigzag GNRs (GNCs). 
}
  \label{fig:zigzag1} 
\end{figure}

As shown in Fig.~\ref{fig:zigzag1} the conductance of a zigzag GNR with a width $W_\mathrm{L}=599\, a_0$ presents a steplike dependence on $E_\mathrm{F}$ (thick black solid line).  The first conductance plateau at $G=G_0$ corresponds to the chiral mode in the direction of the current,  while the following plateaus are separated by $2G_0$,  as a consequence of the mode degeneracy arising from the two graphene valleys.  

Once a constriction is defined on a zigzag GNR (with its axis rotated an angle of $\pi/2$ with respect to the setup of Fig.~\ref{fig:lattice}),  the resulting conductance is reduced as compared to that of the GNR (thick colored lines in Fig.~\ref{fig:zigzag1}).  As remarked in Ref.~\cite{Ihnatsenka:2012},  the conductance plateaus for zigzag edges are better defined than in the armchair case,  especially for the case of a wide GNC.  At low energies,  the GNC with the largest width ($W_\mathrm{C}=499\, a_0$,  brown solid line) does not break the perfectly conducting channel of the circumjacent GNR.  Narrower GNCs (dark blue,  orange,  green,  and violet solid thick curves) destroy the perfect conducting channel at low energies, where similarly to the case of armchair GNCs of Fig.~\ref{fig:TvsE}, sharp resonances associated with quasibound states can be observed \cite{YJXiong:2011,Deng2014}.  Thin dashed lines represent the conductance of a GNR with a width equal to the narrowest distance $W_\mathrm{C}$ of the GNC with the corresponding color.  In the zigzag case the conductance of a wide GNC can be larger than the one of the corresponding GNR of width $W_\mathrm{C}$.  In Fig.~\ref{fig:zigzag1} the thick brown and dark blue solid lines may go,  in some energy intervals,  above the corresponding dashed lines whenever the fictitious GNR has its conductance set at $G_0$ by the perfect conducting channel,  while new transverse modes are open in the GNC setup. 

The slightly better quality of the conductance plateaus for zigzag edges,  as compared with the armchair case,  can be understood from the transmission eigenmode decomposition of the conductance.  In the former case the modes are turned on as $E_\mathrm{F}$ increases (not shown), attaining unitary transmission more sharply than in the case of armchair edges presented in Fig.~\ref{fig:TvsEmd}. 

\begin{figure}[tb]
  \includegraphics[width=\linewidth]{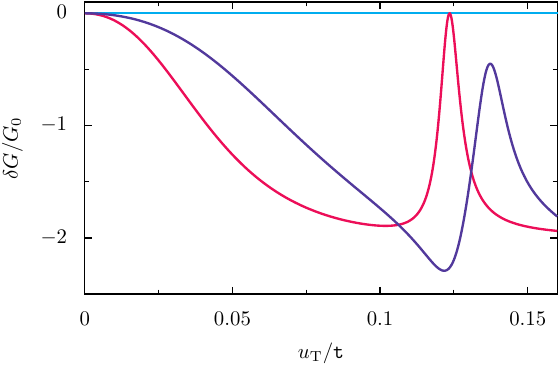} 
\caption{SGM conductance correction for a zigzag GNR of width $W_\mathrm{L}=599\, a_0$,  as a function of the tip-potential strength, when a tip with $d=20\, a_0$ is placed on the axis of the strip,  for three cases where the unperturbed conductance is in the first (blue solid line), second (red solid line), and third (purple solid line) plateau (points $R_1$, $R_2$, and $R_3$ of Fig.~\ref{fig:zigzag1},  respectively).}
  \label{fig:zigzag2} 
\end{figure}

The response of an SGM tip for a zigzag GNR presents some similarities and differences with respect to the armchair case.  In Fig.~\ref{fig:zigzag2} we show the numerically obtained SGM corrections as a function of $u_\mathrm{T}$,  for a tip with $d=20\, a_0$ placed on the longitudinal axis of the strip ($y=W/2$) for three cases where the unperturbed conductance is in the first (blue solid line), second (red solid line), and third (purple solid line) plateau (points $R_1$,  $R_2$,  and $R_3$ of Fig.~\ref{fig:zigzag1},  respectively).  Similarly to the case of an armchair GNR presented in Fig.~\ref{fig:SGMfV}(a),  the perfectly conducting channel seems to be unaffected by the tip (blue line),  while the other two unperturbed conditions exhibit an initial quadratic dependence on $u_\mathrm{T}$ followed by a revival of the conductance for larger potential strengths (red and purple lines corresponding to $R_2$ and $R_3$,  respectively).  However,  this simple picture is modified when the perfect conductance of the chiral modes is destroyed by the effect of the tip. 

\begin{figure}[tb]
  \includegraphics[width=\linewidth]{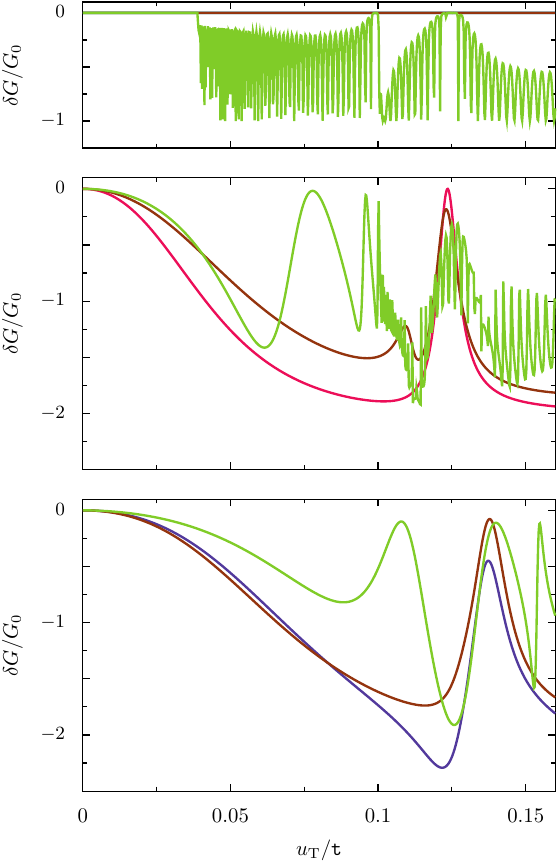} 
\caption{SGM conductance correction for a zigzag GNR of width $W_\mathrm{L}=599\, a_0$,  as a function of the tip-potential strength,  when a tip with $d=20\, a_0$ is placed at the center of the strip (same color convention as in Fig.~\ref{fig:zigzag2}),  at $y=W/4$ (brown),  and at $y=W/8$ (green). Top, middle, and bottom panels correspond,  respectively,  to the three unperturbed conditions defined by the points $R_1$,  $R_2$,  and $R_3$ defined in Fig.~\ref{fig:zigzag1}. }
  \label{fig:zigzag3} 
\end{figure}

In Fig.~\ref{fig:zigzag3} we present the conductance correction as a function of $u_\mathrm{T}$ for $d=20\, a_0$ and the unperturbed conditions defined by the points $R_1$, $R_2$, and $R_3$ of Fig.~\ref{fig:zigzag1} (top, middle, and bottom panels,  respectively) when the tip is placed at the center of the strip (same color convention as in 
Fig.~\ref{fig:zigzag2}),  at $y=W/4$ (brown),  and at $y=W/8$ (green).  Remarkable fluctuations appear when the tip approaches the edges,  an effect already noticed in the numerical simulations of Ref.~\cite{Mrenca_2015}.  Similar features can be observed if the tip is kept on the axis of the strip,  while its extent $d$ is increased.  For $d=60\, a_0$ the perfect conductance is lost for $u_\mathrm{T} \simeq 0.075 \mathtt{t}$,  and strong oscillations set in afterwards (not shown).  These results are consistent with the finding of Ref.~\cite{Akhmerov2008} evidencing that,  even a smooth potential,  if it is strong enough to create a local $p$-$n$ junction,  leads to considerable scattering in a zigzag GNR,  since both valleys are connected by the edge state.  

The perturbative approach to the SGM response developed in Sec.\ \ref{sec:perturbative} is difficult to generalize to zigzag edges,  due to peculiarities of the GNR spectrum.  The existence of flat quasidegenerate bands prevents us from performing the energy integrations as in Eqs.~\eqref{eq:energyintegrations} (upon which the perturbative approach is based),  since it is not correct to restrict the energy interval of integration to that where $\varepsilon$ belongs.  Moreover,  the finite size of the zigzag Brillouin zone cannot be represented by the Dirac equation,  as this continuous description does not account for the differences that  arise according to the parity of the number of atoms $M$ across the transverse direction,  i.e., the contrast between GNRs having the longitudinal axis of symmetry (zigzag configuration, even $M$) and without it (antizigzag configuration,  odd $M$) \cite{Akhmerov2008}.  

The tight-binding model allows one to obtain the form of the edge-state wave functions \cite{Zarea_2009,Wakabayashi2010,Wakabayashi2012,Orlof2013,Talkachov2022}, 
and thus the intervalley tip potential matrix element.  Proceeding as in Ref.~\cite{Akhmerov2008} by restricting the Hilbert space to the lowest $|\varepsilon|$ states permits to understand the observed dependence of the SGM response on the tip position or strength,  as well as the relevance of the parity of $M$.  The dependence on the parity of the number of atoms across the transverse direction appears in the SGM results of zigzag GNRs.  The $u_\mathrm{T}$ dependence of the conductance correction at different points of the transverse cross section for $W_\mathrm{L}=598\, a_0$ (not shown) presents qualitative differences with respect to that of $W_\mathrm{L}=599\, a_0$ shown in Fig.~\ref{fig:zigzag3}.  And related departures appear when $d$ is varied for two strip widths corresponding to different parities.

\begin{figure}[tb]
  \includegraphics[width=\linewidth]{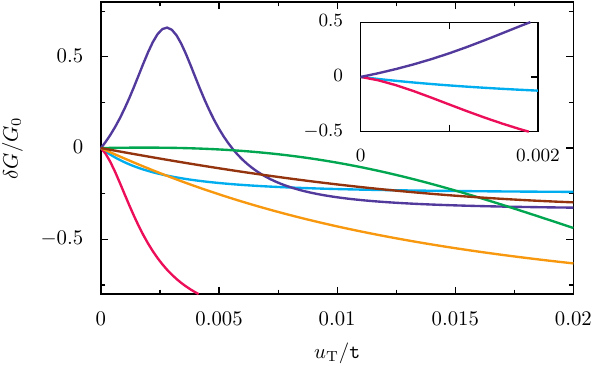} 
  \caption{SGM conductance correction for a GNC defined in a zigzag GNR with $W_\mathrm{C}=199\, a_0$ and $C=\unit[25]{nm}$ as a function of the tip strength,  for a tip of width $d= 20 \, a_0$ placed at the center of the constriction.  The different solid lines correspond to the unperturbed conditions characterized by the points $T_i$ of the same color in Fig.~\ref{fig:zigzag1}.  Inset: Detail of the small $u_\mathrm{T}$ region for the conditions defined by the points $T_i$,  with $i=1,2,3$ characterizing a peak of the unperturbed conductance.}
  \label{fig:zigzag4} 
\end{figure}

The SGM of a GNC defined on a zigzag GNR exhibits similar features as in the case of armchair edges.  In Fig.~\ref{fig:zigzag4} we present the $u_\mathrm{T}$ dependence of the conductance correction for a tip with $d= 20 \, a_0$ placed at the center of the constriction for the unperturbed conditions defined by the points $T_i$ indicated in Fig.~\ref{fig:zigzag1}.  Similarly to the results of Fig.\ \ref{fig:SGMQPC},  the conductance corrections present an initial $u_\mathrm{T}$ dependence which is approximately linear for most of the unperturbed conditions (i.e., points $T_1$,   $T_3$,  $T_4$, and $T_6$),  turning into a quadratic one for the conditions corresponding to a conductance maximum (point $T_2$) or an approximate conductance plateau (point $T_5$). 

\begin{figure}[tb]
  \includegraphics[width=\linewidth]{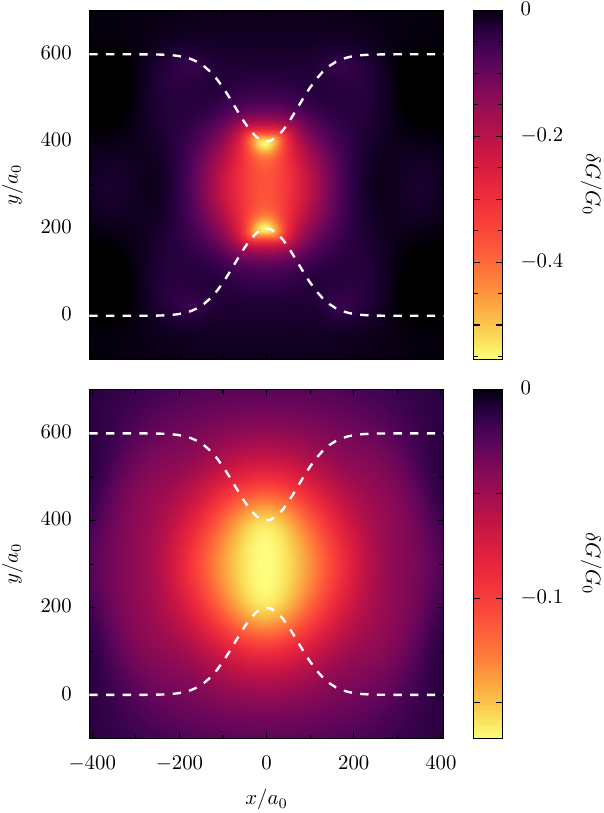}
     \caption{SGM scan for a GNC defined in a zigzag GNR,  with $W_\mathrm{L}=599\, a_0$, $W_\mathrm{C}=199\, a_0$, and $C=\unit[25]{nm}$ (white dashed lines),  for the unperturbed condition corresponding to the point $T_4$ in Fig.~\ref{fig:zigzag1}. The tip potential strength is $u_{\rm T}=0.002\, \mathtt{t}$, while the tip extent is $d=20\, a_0$ ($d=100\, a_0$) for the upper (lower) panel.} 
  \label{fig:zigzag5} 
\end{figure}

The SGM scan of Fig.~\ref{fig:zigzag5} is similar to that of Fig.\ \ref{fig:SGMscan_QPC} (up to the $\pi/2$ rotation).  The spatial feature of two local minima close to the edges in the narrowest part of the GNC appearing for small tips ($d=20\, a_0$,  upper panel),  is blurred for larger tip strengths ($d=100\, a_0$,  lower panel).  There are no oscillations of the SGM response close to the edges as in the case of GNRs,  since the GNC effectively breaks the perfectly conducting channel.


\section{Conclusions}
\label{sec:conclusions}
We have investigated transport in graphene nanoribbons under the influence of the potential of an SGM tip. We have extended the perturbative theory for the SGM response in the regime of noninvasive tips to the case of graphene, and we have demonstrated numerically the validity of the results in the limit of weak tips. On the conductance plateaus observed in metallic armchair ribbons, the tip-induced conductance correction is of second order in the tip strength, and always negative.
A particular situation occurs in the zero-transverse-energy mode, where the second-order correction vanishes, leading to nearly perfect transmission. A data collapse appears when the conductance correction is plotted as a particular combination of the tip strength and size, indicating that a small and strong tip leads to the same lowest-order correction as a larger and correspondingly weaker tip.   

For the regime of stronger tips, we have found that the conductance contributions from channels above the lowest one exhibit a revival to full transmission at tip strengths that decrease with the tip size. These conductance resonances are interpreted in terms of confined states under the tip potential.

The theory has been applied to constrictions defined in a nanoribbon. In this case,  the conductance plateaus are typically lost,  and the perturbation approach predicts,  for the regime of noninvasive tips,  the dominance of a conductance correction that is linear in tip strength. Numerical results confirm such a conclusion,  and a quadratic dependence is obtained in the faint surviving plateaus,  as well as when the unperturbed condition corresponds to a maximum of the conductance. 

Numerical studies of the conductance corrections as a function of tip position yield spatial features when the tip is placed at the narrowest part of the constriction with local maxima close to the edges.  Such a behavior is related to a concentration of the unperturbed PLDOS at those points. Nevertheless, the current density does not exhibit a related behavior. 
 
Zigzag nanoribbons present a similar SGM response as in the armchair case,  with quadratic conductance corrections,  except when the tip potential close to the borders is strong enough to create local $n$-$p$ junctions and destroy the perfect conductance of the chiral edge states.  A GNC defined on a zigzag nanoribbon typically breaks the  perfect channel transmission,  and thus the action of the tip is similar to that observed for armchair edges.  This result is important, as it is for armchair edges that the perturbative approach for noninvasive tips has been developed, moreover since a GNR with an orientation intermediate between zigzag and armchair has been shown to effectively behave as having zigzag edges \cite{1996-PRB-Nakada}.

Several of the previously mentioned theoretical findings can be experimentally checked in graphene nanostructures,  i.e., the initial linear versus quadratic dependence of the SGM correction on the tip-potential strength and its universal scaling,  the revival of perfect conductance outside the noninvasive regime,  the spatial features of the SGM scans in nanoconstrictions,  and the possible destruction of the perfectly conducting channels.  

The extension of our theoretical approach to micrometer sizes could allow to analyze the results of  Ref.~\cite{brun2019} and make the connection with the semiclassical approaches to electron optics in graphene \cite{Cserti2007,Paredes2021}, as well as consider the 
Ohmic-to-ballistic transition studied in  Ref.~\cite{Jenkins2022}.  
Moreover, the perturbative expansion developed in our work can be extended to the magnetic field case, in order to address the 
diversity of field-dependent SGM experiments \cite{Morikawa2015,Bhandari2016,Bours2017,Berezovsky2010b,Chuang2016,Xiang2016,Moreau2021,brun2020,Moreau2021b}.
Other possible generalizations of the theory concern bilayer graphene,  where SGM has allowed detection of localized states  in narrow channels \cite{Gold2020} and the observation of electronic jets emanating from a constriction \cite{Gold2021}, 
as well as transition metal dichalcogenide nanostructures \cite{Bhandari2018,Prokop2020}.

\section*{Acknowledgments}

We thank Eros Mariani for useful discussions.
This work was supported by the National Natural Science Foundation of China under Grant No.\ 12047501. X.C.\ acknowledges the financial support from the China Scholarship Council No.\ 202006180040 and the Programme Doctoral International of the University of Strasbourg.

\appendix

\section{Lead and scattering states for an armchair GNR}
\label{sec:appendixA}

In this Appendix we recall the properties of the electronic eigenstates for an armchair nanoribbon and we define the scattering states to be used in the scattering approach for the conductance of a graphene nanostructure connected to semi-infinite leads of the armchair type.  Our focus on the armchair case stems from the availability of analytical expressions that can be readily employed in the perturbative treatment of the SGM response. 

Instead of using the standard continuous form of the wave functions \cite{Brey2006,Beenakker2006,Wurm2009,Bergvall2014},  we adopt a mixed description with a discretization in the direction transverse to the nanoribbon axis,  where each lattice point describes a conventional cell.  This formulation allows to fix the total number of states participating in the perturbative expansion developed in this work.  For an armchair nanoribbon directed along the $y$ direction and with $M$ unit cells along the transverse direction $x$ (see Fig.~\ref{fig:lattice}),  we take the eigenstate basis
\begin{subequations}
\label{eq:armchairnn}
\begin{align}
\label{eq:armchairnntotal}
&  \psi_{s,\varepsilon,a}(m,y) = 
    \frac{1}{\sqrt{2\pi\hbar v_{\rm F}}} \ \mathrm{e}^{{\rm i}sk_ay} \,
    \Phi_{s,\varepsilon,a}(m)
 \, ,
\\
\label{eq:armchairnntrans}
& \Phi_{s,\varepsilon,a}(m) = 
\frac{1}{\sqrt{M+1}}
\left( \frac{|\xi|}{k_a}\right)^{1/2} \sin\left( \frac{\pi a m }{M+1}\right) \,
\mathcal{Z}_{s,\varepsilon,a}
 \, ,
\end{align}
\end{subequations}
with the pseudospinor
\begin{equation}
\label{eq:armchairps}
\mathcal{Z}_{s,\varepsilon,a} = 
\begin{pmatrix}
       (q_a-{\rm i}sk_a)/\xi \\
        1 \\
    \end{pmatrix}
 \, ,
\end{equation}
for $-\infty < y < \infty$, and $m=0,1,\dots,M+1$.  While the actual width of the nanoribbon is $W_\mathrm{L} = Ma_0$,  the inclusion of the fictitious sites $m=0$ and $m=M+1$ where the wave function vanishes (marked as crosses in Fig.~\ref{fig:lattice}) translates into an effective width $W=(M+1)a_0$. The quantum numbers characterizing the basis are the energy $\varepsilon$ ($-\infty < \varepsilon < \infty$), the transverse channel number $a$ (limited by the fact that there cannot be more than $M$ channels), and the direction $s$ ($s=\pm1$) of the longitudinal wave vector $s k_a$ (we choose $k_a \geqslant 0$). In Eq.~\eqref{eq:armchairnntrans} we have defined the transverse wave vector $q_a=\pi a/W - |\mathbf{K}|$ and the scaled energy $\xi = \varepsilon/\hbar v_{\rm F}= \lambda \sqrt{k_a^2+q_a^2}$, where $\lambda = {\rm sgn}(\varepsilon)$. The decomposition \eqref{eq:armchairnntotal} into longitudinal and transverse components is useful for notation purposes and in view of the general relationship \cite{Wurm2011,Carmier2011}
\begin{equation}
\label{eq:gr}
\sum_{m=0}^{M+1} \Phi_{s,\varepsilon,a}^{\dagger}(m) \ \sigma_y \ \Phi_{s',\varepsilon,a'}(m) = \lambda \, s \, \delta_{s,s'} \, \delta_{a,a'} \,  ,
\end{equation}
where $\sigma_y$ is the usual (second) Pauli matrix. 

The orthonormality and completeness conditions for the eigenbasis \eqref{eq:armchairnn} are expressed,  respectively,  as
\begin{subequations}
\label{eq:orthoandcompl}
\begin{align}
\label{eq:ortho}
\sum_{m=0}^{M+1} \int_{-\infty}^{\infty} {\rm d}y\, \psi_{s,\varepsilon,a}^{\dagger}(m,y) \psi_{s^{\prime},\varepsilon^{\prime},a^{\prime}}(m,y)
        = \delta_{s,s^{\prime}}\delta_{a,a^{\prime}}\delta(\varepsilon-\varepsilon^{\prime})
 \, ,
\\
\label{eq:compl}
\int_{-\infty}^{\infty} {\rm d}\varepsilon \sum_{s,a} 
\psi_{s,\varepsilon,a}^{(j)*}(m,y)\psi_{s,\varepsilon,a}^{(j')}(m',y')
        = \delta_{j,j'}\delta_{m,m^{\prime}}\delta(y-y^{\prime})
 \, ,
\end{align}
\end{subequations}
where the indices $j$ and $j'$ label the pseudospinor components.  Using the relationship \eqref{eq:gr} we obtain the electrical current per unit energy associated with the state $(s,\varepsilon,a)$ as
\begin{equation}
I_{s,\varepsilon,a} = 
e v_{\rm F} \sum_{m=0}^{M+1} \psi_{s,\varepsilon,a}^{\dagger}(m,y) \ \sigma_y \ \psi_{s,\varepsilon,a}(m,y) = \lambda \ s \ \frac{e}{2\pi \hbar} \,  .
\end{equation}

Two different cases can be distinguished among armchair nanoribbons,  depending on whether or not $M+1$ is a multiple of $3$.  The first case is that of metallic nanoribbons with a zero-transverse-energy mode for $a^{*}=4(M+1)/3$ and $\varepsilon = \lambda \hbar v_{\rm F} k_{a^{*}}$,  which is nondegenerate,  while all the other modes are doubly degenerate [i.e., for ${\bar a}=8(M+1)/3-a$ we have $q_{\bar a} = q_a$].  The second case is that of semiconducting nanoribbons with an energy gap around $\varepsilon = 0$, and nondegenerate modes everywhere in the spectrum.  

The scattering approach to quantum transport in graphene can be developed along similar lines as for the 2DEG of semiconductor-based heterojunctions \cite{jalabert2016},  from the nanoribbon eigenstates \eqref{eq:armchairnn},  by defining the incoming ($-$) and outgoing ($+$) modes (lead states)
\begin{equation}
\label{eq:modes}
    \varphi_{l,\varepsilon,a}^{(\mp)}(m,y) = 
    \frac{1}{\sqrt{2\pi\hbar v_{\rm F}}} \ \mathrm{e}^{{\rm i}  s  k_a^{\mp} y} \
    \Phi_{s,\varepsilon,a}(m)
\end{equation}
in the semi-infinite leads.
We note $l=1$ ($2$) for the lower (upper) lead describing $y<0$ ($y>0$), cf.\ Fig.~\ref{fig:lattice}. The direction $s$ of the longitudinal wave vector is defined such that $\lambda \, s =1$ for the up movers ($\varphi_{1,\varepsilon,a}^{(-)}$ and $\varphi_{2,\varepsilon,a}^{(+)}$) and $\lambda \, s =-1$ for the down movers ($\varphi_{2,\varepsilon,a}^{(-)}$ and $\varphi_{1,\varepsilon,a}^{(+)}$). An infinitesimal imaginary part is given to $k_a$ in order to define the proper time ordering, and thus we note $k_a^{\mp}=k_a \mp {\rm i} \lambda \eta$ (with $\eta \to 0^+$). 

Once a quantum-coherent scatterer (of linear extension $L$ in the $y$ direction) is placed at the coordinate origin, the incoming modes $\varphi_{1(2),\varepsilon,a}^{(-)}$ give rise to outgoing scattering states, that in the asymptotic regions are given by
\begin{widetext}
\begin{subequations}
\label{eq:scatteringstates}
\begin{align}
\label{eq:scatteringstates1}
\Psi_{1,\varepsilon,a}(m,y)=
\begin{cases}
        \varphi_{1,\varepsilon,a}^{(-)}(m,y)+\sum_{b}r_{ba}\varphi_{1,\varepsilon,b}^{(+)}(m,y), & y<0,\\
        \sum_{b}t_{ba}\varphi_{2,\varepsilon,b}^{(+)}(m,y),& y>0,
\end{cases}        
\\
\label{eq:scatteringstates2}
\Psi_{2,\varepsilon,a}(m,y)=
\begin{cases}
        \sum_{b}t_{ba}^{\prime}\varphi_{1,\varepsilon,b}^{(+)}(m,y),& y<0,\\
        \varphi_{2,\varepsilon,a}^{(-)}(m,y)+\sum_{b}r_{ba}^{\prime}\varphi_{2,\varepsilon,b}^{(+)}(m,y),& y>0 .
\end{cases}
\end{align}
\end{subequations}
The sums are carried over the number of propagating modes $N$,  which is the dimension of the matrices $r$ ($r'$) and $t$ ($t'$) characterizing the reflection and transmission amplitudes from lead $l=1$ ($l=2$).  We do not explicitly indicate the energy dependence of the scattering amplitudes. 

The outgoing scattering states constitute an eigenbasis. Therefore,  the retarded Green function is a $2 \times 2$ matrix admitting the spectral decomposition
\begin{equation}
    \label{Green_function}
\mathcal{G}(m,y;m^{\prime},y^{\prime};\varepsilon)=\sum_{\barl=1}^{2} \int_{-\infty}^{\infty}
{\rm d}\barep \sum_{\bara} \frac{\Psi_{\barl,\barep,\bara}(m,y)\Psi_{\barl,\barep,\bara}^{\dagger}(m^{\prime},y^{\prime})}{\varepsilon^{+}-\barep} \,  ,
\end{equation}
with $\varepsilon^{+} = \varepsilon + {\rm i}\eta$.  Taking $y'<0$ and $y>0$ we consider
\begin{align}
\label{eq:FLderivation1}
& \sum_{m,m'=0}^{M+1} \
\varphi_{2,\varepsilon,b}^{(+)\dagger}(m,y) \ \sigma_y \
\mathcal{G}(m,y;m^{\prime},y^{\prime};\varepsilon) \ \sigma_y \
\varphi_{1,\varepsilon,a}^{(-)}(m',y')  = \nonumber
\\
&\qquad \sum_{m,m'=0}^{M+1} \ \int_{-\infty}^{\infty}
\frac{\dif \barep}{\varepsilon^{+}-\barep} \
\varphi_{2,\varepsilon,b}^{(+)\dagger}(m,y) \ \sigma_y \
\sum_{c}
\left\{
t_{bc} \ \varphi_{2,\barep,b}^{(+)}(m,y)
\left(
\varphi_{1,\barep,c}^{(-)\dagger}(m',y') \  \delta_{c,a} + r_{ac}^{*} \
\varphi_{1,\barep,a}^{(+)\dagger}(m',y')
\right)  \right.  \nonumber
\\
&
\hspace{15ex}+
\left.
\left(
\varphi_{2,\barep,c}^{(-)}(m,y) \  \delta_{b,c} + r_{bc}^{\prime} \
\varphi_{2,\barep,b}^{(+)}(m,y)
\right)
t_{ac}^{\prime *} \ \varphi_{1,\barep,a}^{(+)\dagger}(m',y')
\right\}
\ \sigma_y \
\varphi_{1,\varepsilon,a}^{(-)}(m',y')
 \, ,
\end{align}
where we have only kept the terms that survive the sums over $m$ and $m'$.  As in Appendix \ref{sec:appendixB},  we assume that the energy integration is dominated by the values of $\barep \simeq \varepsilon$,  and therefore we restrict the integration as to have $ {\bar \lambda}  = \lambda$ (and therefore ${\bar s} = s$). The $\barep$ integral can be done by performing the change of variables from $\barep$ to ${\bar k}_a$,  or to ${\bar k}_b$,  by using ${\bar \xi} = \barep/\hbar v_{\rm F}= {\bar \lambda}({\bar k}_a^2+q_a^2)^{1/2}={\bar \lambda}({\bar k}_b^2+q_b^2)^{1/2}$ and $\xi = \varepsilon/\hbar v_{\rm F}= \lambda({k}_a^2+q_a^2)^{1/2}=\lambda({k}_b^2+q_b^2)^{1/2}$. Implementing the latter change for the first term of the curly bracket and the former change for the second term,  and using Eq.~\eqref{eq:energyintegrations},  we can write expression \eqref{eq:FLderivation1} as
\begin{equation}
- \frac{\rm i}{\hbar \vf} \ t_{ba} \sum_{m'=0}^{M+1}
\varphi_{1,\varepsilon,a}^{(-)\dagger}(m',y') \ \sigma_y \
\varphi_{1,\varepsilon,a}^{(-)}(m',y') \, .
\end{equation}
From definition \eqref{eq:modes} of the incoming modes and the  general relationship \eqref{eq:gr} for the case $\lambda s = 1$,  we
obtain the Fisher-Lee relation for graphene \cite{Wurm2011,Carmier2011}
\begin{equation}
    \label{Fisher_Lee_graphene}
    t_{ba}=2\pi{\rm i}(\hbar \vf)^2\sum_{m,m^{\prime}}
\varphi_{2,\varepsilon,b}^{(+)\dagger}(m,y) \ \sigma_y \ \mathcal{G}(m,y;m^{\prime},y^{\prime};\varepsilon)\ \sigma_y \
    \varphi_{1,\varepsilon,a}^{(-)}(m^{\prime},y^{\prime}) \,  ,
\end{equation}
which, together with Eq.~\eqref{eq:conductance}, gives access to the electron conductance from the knowledge of the Green function,  and it is then used in the numerical and analytical approaches of this work.

\section{Matrix elements of the tip potential for armchair GNRs}
\label{sec:appendixC}

For an armchair GNR,  according to definition \eqref{eq:meotp},  the matrix element of the tip potential between two scattering states impinging from opposite sides with the same energy $\varepsilon$   can be written as
\begin{equation}
\label{eq:meotp2}
\mathcal{U}_{a,a^{\prime}}^{2,1}(\varepsilon,\varepsilon) =  \frac{1}{2  \pi \hbar \vf} \
\sum_{m=0}^{M+1} \int_{-\infty}^{\infty} \dif y \
\mathrm{e}^{{\rm i} \lambda \left(k_a+k_{a^{\prime}}\right)y} \
 \UT(m a_0,y) \ 
 \Phi_{-\lambda,\varepsilon,a}^{\dagger}(m) \ 
 \Phi_{\lambda,\varepsilon,a'}(m) 
 \,  ,
\end{equation}
where we have used that $s' = -s = - \lambda$.  According to Eqs.~\eqref{eq:armchairnntrans} and \eqref{eq:armchairps},  we have
\begin{equation}
 \Phi_{-\lambda,\varepsilon,a}^{\dagger}(m) \
 \Phi_{\lambda,\varepsilon,a'}(m) = \frac{1}{M+1} \
 \frac{|\xi|}{\left(k_a k_{a'} \right)^{1/2}} \
\sin\left( \frac{\pi a m }{M+1}\right) \
\sin\left( \frac{\pi a' m }{M+1}\right) \
\mathcal{Z}_{-\lambda,\varepsilon,a}^{\dagger} \
\mathcal{Z}_{\lambda,\varepsilon,a'}
\end{equation}
and
\begin{equation}
\mathcal{Z}_{-\lambda,\varepsilon,a}^{\dagger} \
\mathcal{Z}_{\lambda,\varepsilon,a'} = 1 +
\frac{1}{ \xi^2}
    \left(q_a-{\ii} \lambda k_a\right)\left(q_{a'}-{\ii} \lambda k_{a'}\right)
 \,  .
\end{equation}

In the case of a metallic GNR, for the zero-transverse-energy mode with $a^{*}= 4(M+1)/3$ and $q_{a^{*}}=0$,  we immediately see that $\mathcal{Z}_{-\lambda,\varepsilon,a^{*}}^{\dagger} \mathcal{Z}_{\lambda,\varepsilon,a^{*}}=0$.  Therefore,  the matrix element $\mathcal{U}_{a^{*},a^{*}}^{2,1}(\varepsilon,\varepsilon)$ vanishes independently of the strength and the position of the tip, and more generally of the features of the perturbing potential, provided that the latter is long ranged \cite{Yamamoto2009} (and as a consequence it does not induce intervalley scattering).

Leaving aside the trivial case of $a=a^{*}$ previously discussed,  we readily perform the $y$ integral in Eq.~\eqref{eq:meotp2}. Moreover, trading the discrete index $m$ by the continuous variable $x=m a_0$,  and noticing that for the lowest-transverse-energy modes the product of the two sines is a rapidly oscillating function of $x$ in comparison with the other terms of the integrand,  we can write
\begin{equation}
\label{eq:meotp3}
\mathcal{U}_{a,a^{\prime}}^{2,1}(\varepsilon,\varepsilon) =  
\frac{\delta_{a,a'}}{4  \pi \hbar \vf} \ \frac{u_{\rm T} \ d}{W} \
\mathrm{e}^{2 {\rm i} \lambda k_a y_{\rm T}} \ \frac{|\xi|}{k_a} \
\left[ 1 +
    \left(\frac{q_a-{\ii} \lambda k_a}{ \xi}\right)^2 \right]
 \int_{0}^{W} \dif x \
 \frac{\exp{\left(-2 k_a d \sqrt{1+(x-x_{\rm T})^2/d^2}\right)}}{\sqrt{1+(x-x_{\rm T})^2/d^2}} \
 \,  .
\end{equation}

For $d \ll W$,  and leaving aside the cases where the distance from the tip to the boundaries is of the order of $d$,  we can push the limits of the $x$ integration to $\mp \infty$,  obtaining
\begin{equation}
\label{eq:meotp4}
\mathcal{U}_{a,a^{\prime}}^{2,1}(\varepsilon,\varepsilon) =  
\frac{\delta_{a,a'}}{2  \pi \hbar \vf} \ \frac{u_{\rm T} \ d^2}{W} \
\mathrm{e}^{2 {\rm i} \lambda k_a y_{\rm T}} \ \frac{|\xi|}{k_a} \
\left[ 1 +
    \left(\frac{q_a-{\ii} \lambda k_a}{ \xi}\right)^2 \right] K_0(2 k_a d)
 \,  ,
\end{equation}
where $K_0$ is the zeroth-order modified Bessel function of the second kind.  

The case of equal-energy scattering states impinging from the same side can be worked out similarly.  For $l=1$ we have 
\begin{equation}
\label{eq:meotp5}
\mathcal{U}_{a,a^{\prime}}^{1,1}(\varepsilon,\varepsilon) =  \frac{1}{2  \pi \hbar \vf} \
\sum_{m=0}^{M+1} \int_{-\infty}^{\infty} \dif y \
\mathrm{e}^{{\rm i} \lambda \left(k_{a^{\prime}}-k_{a}\right)y} \
 \UT(m a_0,y) \ 
 \Phi_{\lambda,\varepsilon,a}^{\dagger}(m) \ 
 \Phi_{\lambda,\varepsilon,a'}(m) 
 \,  .
\end{equation}
As in the previous case,  we take into account that the tip potential is smooth on the scale of the lattice constant,  we readily perform the $y$ integral,  and we convert the $m$ sum into an $x$ integral,  obtaining 
\begin{equation}
\label{eq:meotp6}
\mathcal{U}_{a,a^{\prime}}^{1,1}(\varepsilon,\varepsilon) =  
\frac{\delta_{a,a'}}{2  \pi \hbar \vf} \ \frac{u_{\rm T} \ d}{W} \
\frac{|\xi|}{k_a} \
 \int_{0}^{W} \dif x \
 \frac{1}{\sqrt{1+(x-x_{\rm T})^2/d^2}} \
 \,  .
\end{equation}
Performing the $x$ integral, we have
\begin{equation}
\label{eq:meotp7}
\mathcal{U}_{a,a^{\prime}}^{1,1}(\varepsilon,\varepsilon) =  
\frac{\delta_{a,a'}}{2  \pi \hbar \vf} \ 
\frac{u_{\rm T} \ d^2}{W} \ \frac{|\xi|}{k_a} \
\ln{\left(
\frac{\sqrt{(W-x_{\rm T})^2/d^2+1}
+(W-x_{\rm T})/{d}}
{\sqrt{(x_{\rm T}/{d})^2+1}-x_{\rm T}/d}
\right)}
 \,  .
\end{equation}
%

\section{Energy integrations for the SGM corrections}
\label{sec:appendixB}

In this Appendix we work out a few integrations appearing in the perturbative treatment of the tip potential.  We first present the results:
\begin{subequations}
\label{eq:energyintegrations}
\begin{align}
\label{eq:energyintegrations1}
& \sum_{m=0}^{M+1} \int_{-\infty}^{\infty} 
\frac{\dif \barep}{\varepsilon^{+}-\barep} \
\varphi_{2,\varepsilon,b}^{(+)\dagger}(m,y) \ \sigma_y \ 
\varphi_{2,\barep,\barb}^{(+)}(m,y) \ f_1(\varepsilon,\barep) =
- \frac{\rm i}{\hbar \vf} \delta_{b,\barb} \ f_1(\varepsilon,\barep)
 \,  ,
\\
\label{eq:energyintegrations2}
& \sum_{m'=0}^{M+1} \int_{-\infty}^{\infty} 
\frac{\dif \barep}{\varepsilon^{+}-\barep} \
\varphi_{1,\barep,\bara}^{(-)\dagger}(m',y') \ \sigma_y \ 
\varphi_{1,\varepsilon,a}^{(-)}(m',y') \ f_2(\varepsilon,\barep) =
- \frac{\rm i}{\hbar \vf} \delta_{a,\bara} \ f_2(\varepsilon,\barep)
 \, ,
\\
\label{eq:energyintegrations3}
& \sum_{m=0}^{M+1} \int_{-\infty}^{\infty} 
\frac{\dif \barep}{\varepsilon^{+}-\barep} \
\varphi_{2,\varepsilon,b}^{(+)\dagger}(m,y) \ \sigma_y \ 
\varphi_{2,\barep,\barb}^{(-)}(m,y) \ f_3(\varepsilon,\barep) = 0
 \,  ,
\\
\label{eq:energyintegrations4}
& \sum_{m'=0}^{M+1} \int_{-\infty}^{\infty} 
\frac{\dif \barep}{\varepsilon^{+}-\barep} \
\varphi_{1,\barep,\bara}^{(+)\dagger}(m',y') \ \sigma_y \ 
\varphi_{1,\varepsilon,a}^{(-)}(m',y') \ f_4(\varepsilon,\barep) = 0
 \, , 
\end{align}
\end{subequations}
for $y>0$ and $y'<0$,  while $f_i(\varepsilon,\barep)$ (with $i=1,\dots,4$) are arbitrary functions assumed to have a smooth dependence on $\varepsilon$ and~$\barep$. 

According to definition \eqref{eq:modes}, in the integration of Eq.~\eqref{eq:energyintegrations1} we have $\lambda \, s = {\bar \lambda} \, {\bar s} = 1$.  The sum over $m$ translates into the restriction $b=\barb$.  Since the energy integration is dominated by the values of $\barep \simeq \varepsilon$,  we can restrict the integration as to have $ {\bar \lambda}  = \lambda$ (and therefore ${\bar s} = s$).  Performing the change of variables from $\barep$ to ${\bar k}_b$ such that ${\bar \xi} = \barep/\hbar v_{\rm F}= {\bar \lambda}({\bar k}_b^2+q_b^2)^{1/2}$,  the left-hand side of Eq.~\eqref{eq:energyintegrations1} can be written as 
\begin{align}
    \label{integration1a}
 &   - \frac{{\rm i} \  \delta_{b,\barb} \lambda}{4 \pi \hbar \vf} \int_{0}^{\infty} \ 
    \dif \bark_b \   \left(\frac{\bark_b}{k_b }\right)^{1/2} \
    \left|\frac{\xi}{{\bar \xi}}\right|^{1/2}
    \left(\frac{q_b-{\ii} sk_b}{\xi}-\frac{q_b+{\ii} s\bark_b}{\bar \xi}\right)
    \left(\sqrt{\bark_b^2+q_b^2} + \sqrt{k_b^2+q_b^2} \right)  \nonumber \\
    &\qquad \times
    \frac{1}{\bark_b-(k_b+{\rm i}\lambda \eta)} \
    \frac{1}{\bark_b+(k_b+{\rm i}\lambda \eta)} \
    \mathrm{e}^{{\rm i} s\left(\bark_b^{(+)}-k_b^{(-)}\right)y} \ f_1(\varepsilon,\barep)
    \, .
\end{align}
The integral can be done by contour integration in the complex $\bark_b$ plane.  In the case $\lambda = s = 1$ ($ \lambda =  s = - 1$) the contour should be closed on the upper (lower) half plane with positive (negative) values of ${\rm Im}\{\bark_b\}$ in order to have vanishing contributions from the vertical and quarter-circle segments for large values of $y$ (see Fig.~\ref{fig:contours}).  Such contours leave aside the other poles at $\bark_b = \pm {\rm i}q_b$,  as well as the branch cuts associated with the square root.  Therefore,  the contour integral is determined by the pole at $k_b \pm {\rm i} \eta$,  and we readily recover the result \eqref{eq:energyintegrations1}.  The integral of Eq.~\eqref{eq:energyintegrations2}, where we also have $\lambda \, s = {\bar \lambda} \, {\bar s} = 1$, follows along the same lines of the previous case. 

\begin{figure}[tb]
\includegraphics[width=.4\linewidth]{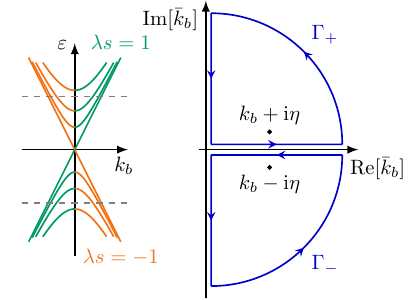}
    \caption{Left: Dispersion relation of the lowest-energy lead states for a metallic armchair GNR showing the up movers ($\lambda \, s =1$) in green and the down movers ($\lambda \, s =-1$) in orange.  The grey dashed horizontal lines of constant energy $\varepsilon$ determine the allowed longitudinal wave vectors $k_b$ of the propagating modes. Right: Contours in the complex $\bark_b$ plane for the integration of Eqs.~\eqref{integration1a} and \eqref{integration3a}.  For the integration of Eq.~\eqref{integration1a} the upper (lower) contour $\Gamma_{+}$ ($\Gamma_{-}$) should be used for the case $\lambda = s = 1$ ($ \lambda =  s = - 1$).  The opposite choice is required for the integration   of Eq.~\eqref{integration3a}.}
     \label{fig:contours}
\end{figure}

In the integrand of  Eq.~\eqref{eq:energyintegrations3} we have $\lambda s = - {\bar \lambda} {\bar s} = 1$ and the sum over $m'$ leads to the condition $b=\barb$.  Restricting the interval of the energy integration such that ${\bar \lambda} = \lambda$ (and therefore ${\bar s} = - s$) and performing the same change of variables specified above,  the left-hand side of Eq.~\eqref{eq:energyintegrations3} can be written as 
\begin{align}
    \label{integration3a}
 &   \frac{{\rm i} \  \delta_{b,\barb} \lambda}{4 \pi \hbar \vf} \int_{0}^{\infty} \ 
 \dif \bark_b \   \left(\frac{\bark_b}{k_b }\right)^{1/2} \
    \left|\frac{\xi}{{\bar \xi}}\right|^{1/2}
    \left(\frac{q_b+{\ii} sk_b}{\xi}-\frac{q_b+{\ii} s\bark_b}{\bar \xi}\right)
    \left(\sqrt{\bark_b^2+q_b^2} + \sqrt{k_b^2+q_b^2} \right) \nonumber \\
    &\qquad \times
    \frac{1}{\bark_b-(k_b+{\rm i}\lambda \eta)} \
    \frac{1}{\bark_b+(k_b+{\rm i}\lambda \eta)} \
    \mathrm{e}^{-{\rm i}s\left(\bark_b^{(-)}+k_b^{(-)}\right)y} \ f_3(\varepsilon,\barep)
    \, .
\end{align}
For the case $\lambda = s = 1$ ($\lambda = s = - 1$) the contour should be closed on the lower (upper) half plane with negative (positive) values of ${\rm Im}\{\bark_b\}$ in order to have vanishing contributions from the vertical and quarter-circle segments (see Fig.~\ref{fig:contours}),  and the absence of poles inside the contour results in a vanishing integral,  as stated in Eq.~\eqref{eq:energyintegrations3}.  The integral of Eq.~\eqref{eq:energyintegrations4},  where we also have $\lambda s = - {\bar \lambda} {\bar s} = 1$,  follows along the same lines of the previous case. 

We continue with the demonstration of an important identity concerning the zero-transverse-energy mode of the metallic armchair GNR:
\begin{equation}
\label{eq:energyintegrationsZTEM}
 \int_{-\infty}^{\infty} 
\frac{\dif \barep}{\varepsilon^{+}-\barep} \
\mathcal{U}_{a^{*},a^{*}}^{1,1}(\varepsilon,\barep) \
\mathcal{U}_{a^{*},a^{*}}^{1,1}(\barep,\varepsilon) 
=
- {\rm i}\pi \left[\mathcal{U}_{a^{*},a^{*}}^{1,1}(\varepsilon,\varepsilon) \right]^2
\,  .
\end{equation}

Concentrating on the $\barep$-dependent terms of the integrand above,  we need to evaluate
\begin{equation}
\label{eq:energyintegrationsZTEM2}
 \int_{-\infty}^{\infty} 
\frac{\dif \barep}{\varepsilon^{+}-\barep} \
\Psi_{1,\barep,a^{*}}(m,y) \  
\Psi_{1,\barep,a^{*}}^{\dagger}(m^{\prime},y^{\prime})
=  \frac{1}{\pi\hbar v_{\rm F}} \ \frac{1}{M+1}
\sin\left( \frac{\pi a^{*} m }{M+1}\right) \
\sin\left( \frac{\pi a^{*} m^{\prime}}{M+1}\right)
 \int_{-\infty}^{\infty} 
\frac{\dif \barep}{\varepsilon^{+}-\barep} \
\mathrm{e}^{{\rm i}{\bar \lambda} \bark_{a^{*}} \left(y-y^{\prime} \right)} 
\,  .
\end{equation}
We note $\barep= \hbar v_{\rm F} {\bar \lambda} \bark_{a^{*}}$ and  $\varepsilon= \hbar v_{\rm F} \lambda k_{a^{*}}$.  The above $\barep$ integral can be readily done by contour integral in the complex $\bark_{a^{*}}$ plane,  and thus
\begin{align}
\label{eq:energyintegrationsZTEM3}
 \int_{-\infty}^{\infty} 
\frac{\dif \barep}{\varepsilon^{+}-\barep} \
\mathcal{U}_{a^{*},a^{*}}^{1,1}(\varepsilon,\barep) \
\mathcal{U}_{a^{*},a^{*}}^{1,1}(\barep,\varepsilon) 
=& -
\frac{2 {\rm i}}{\hbar v_{\rm F}} \ \frac{1}{M+1}
\sum_{m=0}^{M+1} \sum_{m'=0}^{M+1}
\sin\left( \frac{\pi a^{*} m }{M+1}\right) \
\sin\left( \frac{\pi a^{*} m^{\prime}}{M+1}\right)
 \int_{-\infty}^{\infty} \dif y \ \int_{-\infty}^{\infty} \dif y^{\prime} 
  \nonumber \\
 &  \times \theta\left(y-y^{\prime}\right) \
\Psi_{1,\varepsilon,a^{*}}^{\dagger}(m,y) \   \UT(m a_0,y) \ 
\mathrm{e}^{{\rm i}\lambda k_{a^{*}} \left(y-y^{\prime} \right)} \
 \UT(m^{\prime} a_0,y^{\prime}) \
\Psi_{1,\varepsilon,a^{*}}(m^{\prime},y^{\prime}) 
\,  ,
\end{align}
where $ \theta$ is the Heaviside step function.  Since the only $(y,y')$ dependence of the integrand is that of $\UT(m a_0,y) \  \UT(m^{\prime} a_0,y)$,  it is easy to see that the $(y,y')$ integrals can be taken as unrestricted,  trading $\theta\left(y-y^{\prime}\right)$ by a factor of $1/2$.  Therefore,  Eq.~\eqref{eq:energyintegrationsZTEM3} can be expressed as the product of two matrix elements of the tip potential,  and we recover the result \eqref{eq:energyintegrationsZTEM},  implying that the principal part of the integral vanishes.

The result \eqref{eq:energyintegrationsZTEM} can be generalized to the case of $n-1$ intermediate energy integrations,  leading to 
\begin{equation}
\label{eq:energyintegrationsZTEMn}
 \int_{-\infty}^{\infty} 
\frac{\dif \barep_2}{\varepsilon^{+}-\barep_2} \
 \int_{-\infty}^{\infty} 
\frac{\dif \barep_3}{\varepsilon^{+}-\barep_3} \
\ldots\
 \int_{-\infty}^{\infty} 
\frac{\dif \barep_n}{\varepsilon^{+}-\barep_n} \
\mathcal{U}_{a^{*},a^{*}}^{1,1}(\varepsilon,\barep_2) \
\mathcal{U}_{a^{*},a^{*}}^{1,1}(\barep_2,\barep_3) \
\ldots\
\mathcal{U}_{a^{*},a^{*}}^{1,1}(\barep_n,\varepsilon)
=
\left[ - {\rm i}\pi\right]^{n-1}  \left[\mathcal{U}_{a^{*},a^{*}}^{1,1}(\varepsilon,\varepsilon) \right]^n
\,  .
\end{equation}
Such a result can be obtained by expressing the matrix elements in terms of the corresponding wave functions,  and then performing the energy integrations independently,  which yields a factor of $\left[ - 2 \pi {\rm i}\right]^{n-1}$ and the constraint $\theta(y_1-y_2) \ \theta(y_2-y_3)  \ldots\ \theta(y_{n-1}-y_n)$.  As before,  the only $(y_1,y_2,\ldots,y_n)$ dependence of the integrand is contained in the tip potentials $\UT$,  and therefore the $(y_1,y_2,\ldots,y_n)$ integral can be taken as unrestricted by trading the Heaviside step function by a factor of $1/2^n$,  which results in the matrix element $\mathcal{U}_{a^{*},a^{*}}^{1,1}(\varepsilon,\varepsilon)$ to the $n$th power.

\end{widetext}

\bibliography{sgm_graphene}

\end{document}